\documentclass[sigconf, nonacm]{acmart}

\usepackage{amsmath}
\usepackage{lmodern}% http://ctan.org/pkg/lm
\usepackage{subcaption}

\usepackage{url}
\def\UrlBreaks{\do\/\do-}
\usepackage{algorithm}
\usepackage{algpseudocode}
\usepackage{xspace}

\makeatletter
\g@addto@macro{\UrlBreaks}{\UrlOrds}
\makeatother

\usepackage{makecell} % Include this in your preamble
\usepackage{stmaryrd}
\usepackage{listings}
\usepackage{color}

\lstdefinestyle{customc}{
  language=C,
  numbers=left,
  stepnumber=1,
  numberstyle=\tiny,
  basicstyle=\ttfamily\small,
  keywordstyle=\color{blue}\bfseries,
  commentstyle=\color{green},
  stringstyle=\color{red},
  frame=none,
  breaklines=true,
  columns=fullflexible
}

\usepackage{tikz}
\usetikzlibrary{shapes,matrix,er,arrows,shadows,trees,positioning}

\usepackage{enumitem}
\setitemize{noitemsep,topsep=0pt,parsep=0pt,partopsep=0pt}

\usepackage[font=small,format=plain,labelfont=bf]{caption}

\DeclareGraphicsExtensions{.png,.pdf}

\def \ShowRealNames {}

\ifx \ShowRealNames \undefined
\newcommand{\companyX}{CompanyX\xspace}
\newcommand{\fullCompanyX}{CompanyX\xspace}

\newcommand{\productX}{ProductX\xspace}
\newcommand{\verA}{VersionA\xspace}
\newcommand{\verB}{VersionB\xspace}
\newcommand{\productY}{ProductY\xspace}
\newcommand{\productZ}{ProductZ\xspace}
\newcommand{\moduleA}{ModuleA\xspace}
\newcommand{\moduleB}{ModuleB\xspace}
\newcommand{\moduleC}{ModuleC\xspace}
\else
\newcommand{\companyX}{VMware\xspace}
\newcommand{\fullCompanyX}{Broadcom Inc.\xspace}

\newcommand{\productX}{vSAN\xspace}
\newcommand{\verA}{OSA\xspace}
\newcommand{\verB}{ESA\xspace}
\newcommand{\productY}{VDFS\xspace}
\newcommand{\productZ}{vSAN File Services\xspace}
\newcommand{\moduleA}{DOM\xspace}
\newcommand{\moduleB}{zDOM\xspace}
\newcommand{\moduleC}{LSOM2\xspace}
\fi

\newcommand{\btree}{B$^+$ tree\xspace}

\begin{document}

\title{Clock2Q+: A Simple and Efficient Replacement Algorithm for Metadata Cache in \companyX \productX}

\author{Yiyan Zhai}
\affiliation{%
  \institution{Carnegie Mellon University}
  \city{Pittsburgh}
  \country{United States}
}
\email{yiyanz@andrew.cmu.edu}

\author{Bintang Dwi Marthen}
\affiliation{%
  \institution{Bandung Institute of Technology}
  \city{Bandung}
  \country{Indonesia}
}
\email{marthen.bintangdwi@gmail.com}

\author{Sarath Balivada}
\affiliation{%
  \institution{\fullCompanyX}
  \city{Bangalore}
  \country{India}
}
\email{sarath.balivada@broadcom.com}

\author{Vamsi Sudhakar Bojji}
\affiliation{%
  \institution{\fullCompanyX}
  \city{Bangalore}
  \country{India}
}
\email{vamsi-sudhakar.bojji@broadcom.com}

\author{Eric Knauft}
\affiliation{%
  \institution{\fullCompanyX}
  \city{Palo Alto}
  \country{United States}
}
\email{eric.knauft@broadcom.com}

\author{Jitender Rohilla}
\affiliation{%
  \institution{\fullCompanyX}
  \city{Palo Alto}
  \country{United States}
}
\email{jitender.rohilla@broadcom.com}

\author{Jiaqi Zuo}
\affiliation{%
  \institution{\fullCompanyX}
  \city{Palo Alto}
  \country{United States}
}
\email{alvin.zuo@broadcom.com}

\author{Quanxing Liu}
\affiliation{%
  \institution{\fullCompanyX}
  \city{Palo Alto}
  \country{United States}
}
\email{quanxing.liu@broadcom.com}

\author{Maxime Austruy}
\affiliation{%
  \institution{\fullCompanyX}
  \city{Pully}
  \country{Switzerland}
}
\email{maxime.austruy@broadcom.com}

\author{Wenguang Wang}
\affiliation{%
  \institution{\fullCompanyX}
  \city{Palo Alto}
  \country{United States}
}
\email{wenguang.wang@broadcom.com}

\author{Juncheng Yang}
\affiliation{%
  \institution{Harvard University}
  \city{Cambridge}
  \country{United States}
}
\email{juncheng@g.harvard.edu}

\begin{abstract}

Cache replacement algorithms are critical building blocks of storage systems. 
This paper examines the characteristics of metadata caches and argues that they inherently exhibit correlated references, even when the corresponding data accesses do not contain correlated references. 
The presence of correlated references reduces the effectiveness of cache replacement algorithms because these references are often mistakenly categorized as hot blocks. 

Clock2Q+ is specifically designed for metadata caches and has been implemented in \productX and \productY, two flagship storage products of \fullCompanyX. 

Similar to S3-FIFO, Clock2Q+ uses three queues; however, Clock2Q+ introduces a correlation window in the Small FIFO queue, where blocks in this window do not set the reference bit. This simple enhancement allows Clock2Q+ to outperform state-of-the-art replacement algorithms. Compared to S3-FIFO, the second-best performing algorithm, Clock2Q+ achieves up to a 28.5\% lower miss ratio on metadata traces. 
Clock2Q+ possesses the essential properties required for large-scale storage systems: it has low CPU overhead on cache hits, low memory overhead, scales efficiently to multiple CPUs, and is both easy to tune and implement. Additionally, Clock2Q+ outperforms state-of-the-art cache replacement algorithms on data traces as well.
\end{abstract}

\maketitle

\section{Introduction}\label{sec:intro}

Storage systems typically consist of multiple tiers, such as CPU caches, memory, and persistent media (e.g., SSDs and HDDs). 
Caches are widely employed to accelerate data access. 
This paper examines block caches (also referred to as page caches, buffer caches, or buffer pools), which are common memory caches that manage fixed-size blocks (e.g., 4KB) to enhance storage access. In a block cache, the fast tier is memory, whereas the slow tier consists of various storage devices, such as SSDs and magnetic drives.

A key challenge in block caching is designing an efficient cache replacement algorithm that determines which data remain in the cache and which data are evicted. An ideal algorithm should aim to achieve:

\begin{itemize}
\item \emph{High hit ratio} across diverse workloads, avoiding pathological cases that degrade the hit ratio.
\item \emph{Low CPU overhead} on cache hits, ensuring that cache access does not become a bottleneck, especially at high hit ratios (close to 100\%).
\item \emph{Scalability to multiple CPUs}. The algorithm should scale efficiently on multi-core systems.
\item \emph{Minimal tuning required} for configuration parameters, enabling good performance across various workloads.
\item \emph{Ease of implementation and debugging}, reducing development effort and ensuring reliability.
\end{itemize}

Block caches store various types of data, which influence the design of the cache replacement algorithm:

\begin{itemize} 
\item \emph{Data Cache}: Stores the actual storage payload (e.g., file data).
\item \emph{Metadata Cache}: Stores metadata used to manage the storage system (e.g., \btree blocks).
\end{itemize}

Many systems maintain separate caches for data and metadata. For example, Linux uses the Buffer Cache to cache file system metadata and the Block Cache to cache file data~\cite{bovet2005linux}. 
\productX, a distributed block storage system, includes both a metadata cache and a data cache. \productX features a small data cache at the \moduleA Client layer and multiple large metadata caches at its \moduleB and \moduleC layers~\cite{clock2q-concurrent-patent, clock2q-patent}.

By analyzing the fundamental properties of data caches and the typical usage of metadata to manage data, this paper highlights key differences between them. In a metadata cache, some blocks may be referenced multiple times within a short period, followed by a long period of inactivity. Such frequent accesses within a short period, referred to as ``correlated references'', should be treated as a single reference by the cache replacement algorithm. If these blocks are misclassified as hot due to the high number of references they receive in this window, they will remain in the cache for an extended period. A block exhibiting this reference pattern is still considered a cold block and should be evicted once its correlated reference period ends. This behavior differs from most data cache access patterns, where cold blocks are typically accessed only once. If this property is not accounted for, a cold block in the metadata cache could be mistakenly treated as a hot block, thereby reducing the overall hit ratio.

Most existing algorithms struggle to handle this access pattern effectively. 
For example, S3-FIFO~\cite{yang_fifo_2023-1} uses a Small FIFO queue to filter out unpopular blocks: if a block does not receive a request while in the Small FIFO queue, it is evicted early. However, this design is vulnerable to correlated accesses because such blocks are misclassified as hot and moved to the Main Clock, even though they will not receive any further requests after the correlation window ends. Therefore, tailoring the cache algorithm specifically for metadata workloads can provide additional performance benefits for the metadata cache.

In this paper, we describe a new cache replacement algorithm, Clock2Q+, that is designed for metadata cache and has been implemented in \productX. 
Clock2Q+ is a hybrid of Clock2Q (\S\ref{subsec:clock2q}) and S3-FIFO~\cite{yang_fifo_2023-1} by combining key design ideas of both algorithms and and improving upon certain aspects of each. 
Clock2Q+ splits the small FIFO queue in S3-FIFO into two ``virtual'' segments, where requests to blocks in the first segment do not set the reference bit. 
Compared to Clock2Q, it reduces cache misses when hot blocks first enter the cache. Compared to S3-FIFO, it reduces cache misses caused by correlated references. 
This simple change allows Clock2Q+ to achieve up to a 28.5\% lower miss ratio than S3-FIFO on metadata cache workloads. Moreover, Clock2Q+ also achieves lower miss ratio on data cache workloads even though they may not exhibit strong correlated access patterns. 
It also provides a low cache hit overhead, high scalability, and remarkable simplicity, which enables quick implementation in production and add other features such as concurrency and resizing. When handling dirty blocks, simple adjustments are made to the algorithm without compromising the cache miss ratio.

This paper has the following contributions:

\begin{itemize}
\item Identifies and establishes that correlated references are a common access pattern in metadata caches.
\item Proposes a simple method to convert data traces into metadata traces. This simplifies the evaluation of cache replacement algorithms on metadata cache workloads as most existing traces are collected from data accesses.
\item Describes Clock2Q+, a scalable, low-overhead cache replacement algorithm implemented in \productX. 
\item Evaluates Clock2Q+ on both metadata and data cache workloads, demonstrating that it outperforms state-of-the-art methods with lower miss ratios. 
\end{itemize}

\section{Metadata Cache and Correlated References}\label{sec:cache-trace-type}

\subsection{Data Cache and Metadata Cache}\label{sec:meta-vs-data}
A data cache is the most common type of cache where the cached data is the actual blocks (from a storage system). Most, if not all, existing caches and replacement algorithms are designed for storing data blocks. 

In addition to data caches, metadata caches (also called index caches in many systems) are also widely used. 
Many storage systems use a form of ordered key-value store, such as \btree, to map the Logical Block Number (LBN) to the Physical Block Number (PBN). 
For example, \productY, \moduleB in \productX \verB, \moduleC in \productX \verA, Linux XFS, and Btrfs all use \btree~\cite{cow-btree-code, xfs, btrfs13}; \moduleC in \productX \verB uses SplinterDB~\cite{splinter-atc-20}; Unix FFS and Linux ext2/ext3 use direct and indirect block maps similar to page table; Linux ext4 uses extent tree~\cite{ext4tree}.
In all these examples, the key-value store has fixed-size leaf blocks that store the LBN to PBN mapping. Because the key-value store is often large, only a small portion of it is cached in the metadata cache discussed in this work.

\subsection{Correlated References}

Storage often consists of multiple layers. Consider a two-layered system where an upper file system, such as ext4, is formatted on top of a disk array box that contains a lower file system. Within this lower file system, there are both a data cache and a metadata cache. The data cache stores common data requested by the upper file system, while the metadata cache stores the disk array's own \btree blocks, which map the logical addresses of data received from the upper file system to physical addresses on disks. The data cache has a lower temporal locality because the upper file system also has a data cache that filters out repeated accesses~\cite{froese96, yadgar_management_2011, yadgar_karma_2007}. 
However, as the following analysis shows, the metadata cache can still experience frequent, short-term accesses to the same metadata blocks, referred to as "correlated references".

\begin{figure}[ht]
  \centering
    \includegraphics[width=0.7\linewidth]{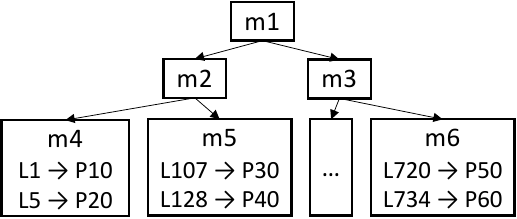}
    \caption{Example \btree for mapping LBN to PBN. Lookup LBN $5$ will read metadata block $m1$, $m2$, $m4$ and find LBN $5$ is mapped to PBN $20$.}
    \label{fig:btree}
\end{figure}

Using \btree as an example, assume the \btree uses LBN as the key and PBN as the value. Figure~\ref{fig:btree} shows some metadata blocks with address $m1$, $m2$, etc., and data blocks with LBN $L1$, $L2$, etc.

If the data trace accesses 4 LBNs in the following order: 1, 5, 107, and 720, the metadata accesses will be: $m1$, $m2$, $m4$, $m1$, $m2$, $m4$, $m1$, $m2$, $m5$, $m1$, $m3$, $m6$. 
It is evident that the non-leaf blocks ($m1$, $m2$, $m3$) are accessed frequently. Any reasonable replacement algorithm will keep non-leaf blocks in the cache. Therefore, when analyzing metadata caches, we ignore non-leaf blocks and focus solely on the leaf blocks, which make up 99\% or more of \btree blocks. 
Excluding non-leaf blocks, the accessed metadata blocks reduce to: $m4$, $m4$, $m5$, $m6$. Since each leaf block contains tuples, $m4$ appears multiple times, even though the upper layer doesn't access any data more than once. 
This is an example of correlated reference -- accessing $m4$ multiple times within a short interval does not mean it is a hot block. This leads to an important observation: even if there are no correlated references in the data cache, its corresponding metadata cache may contain correlated references that must be properly handled by the cache replacement algorithm. 

While we use \btree as an example to analyze access patterns of the metadata cache, the correlated reference access pattern is common across most metadata structures.
A typical metadata block contains multiple smaller entities, such as key-value tuples, bits in a bitmap, or inodes in an inode block. This makes metadata blocks more susceptible to correlated references than data blocks: even if the upper layer visits different entities only once, the metadata block containing these entities is accessed multiple times, leading to correlated references. Thus, we argue that having correlated references is an inherent trait of metadata caches.

\subsection{Deriving Metadata Traces}\label{sec:cache-trace-type:convert}

Trace-driven simulation is a crucial step in evaluating cache replacement algorithms. One challenge in improving replacement algorithms for metadata caches is the lack of available metadata traces. It is straightforward to collect data traces, as they capture the requests that go into storage systems. However, collecting metadata traces is much harder, since the storage system itself has to be instrumented to record each access to metadata blocks. Moreover, each access to a data block may trigger multiple accesses to internal metadata blocks, resulting in much longer metadata traces. Therefore, despite the recent release of multiple open-source cache datasets, all of them are limited to data traces.

To address the lack of metadata traces, we propose a method to derive metadata traces from any existing data traces.
Using Figure~\ref{fig:btree} as an example, if each \btree leaf block stores 100 tuples mapping LBN to PBN, we can estimate that LBN 0-99 is in leaf block $m0$, LBN 100-199 are in leaf block $m1$, and so on. 
With this mapping, we can convert an LBN in the data trace to a metadata block by dividing it by the fan-out: $\ \frac {LBN}{FanOut}$. 
For example, a data cache trace with requests for 1, 5, 107, and 720 would be converted into a metadata trace of 0, 0, 1, and 7.
This approach allows us to generate metadata traces from existing data traces.

\section{Clock2Q+ Design}\label{sec:design}

Given the similarities between the four cache replacement algorithms: 2Q~\cite{johnson_2q_1994}, Clock2Q (the previous cache replacement algorithm in \productX), S3-FIFO~\cite{yang_fifo_2023-1}, and Clock2Q+, we first present and analyze the first three algorithms before introducing the design of Clock2Q+.

\subsection{2Q}\label{subsec:2q}

\begin{figure}[t]
    \centering
      \includegraphics[width=.775\linewidth]{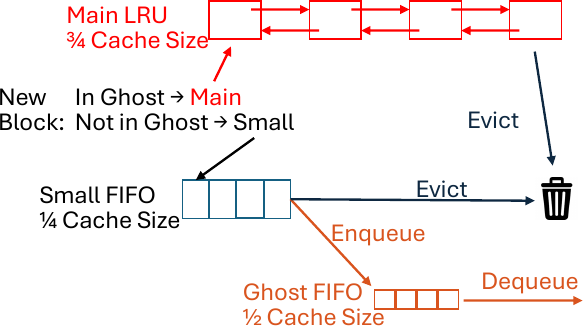}
    \caption{Illustration of 2Q cache replacement algorithm. \vspace{-0.8em}}\label{fig:2q}
\end{figure}

The 2Q~\cite{johnson_2q_1994} algorithm, illustrated in Figure~\ref{fig:2q}, divides the cache into two parts: a Main LRU of $\frac{3}{4}$ of the cache size, which manages blocks using LRU, and a Small FIFO of $\frac{1}{4}$ of the cache size, which manages blocks using a FIFO queue. 
There is also a Ghost FIFO that stores a number of block entries equal to $\frac{1}{2}$ of the cache capacity. The Ghost FIFO only stores block numbers rather than actual data, so its memory usage is small compared to the cache size.

When a new block is accessed, if it is not found in the Ghost FIFO, it enters
the Small FIFO. If the Small FIFO is full, its last block is evicted, and its
block number is added to the Ghost FIFO, which may in turn evict the last block
number from the Ghost FIFO if it is also full.

If a block number is found in the Ghost FIFO, it means that the block is actually hot but was mistakenly evicted from the Small FIFO previously. In this case, the block is added to the Main LRU. Following the LRU eviction policy, it may evict a block from the tail of the LRU. The Ghost FIFO provides the cache with a form of ``long-term memory'' of recently evicted blocks, which is particularly useful when workloads exhibit large loops in access patterns.

If an existing block in the Small FIFO is accessed, no action is needed. If the existing block is accessed in the Main LRU, it follows the standard LRU policy: it is removed from the LRU list and re-inserted at the head of the LRU list.

The Small FIFO serves the purpose of removing correlated references, as such
references typically occur while the blocks are still in the Small FIFO, and
they do not have a chance to be promoted to the Main LRU.

There are several drawbacks to 2Q:

\begin{itemize}
\item \emph{High LRU overhead}. The Main LRU has high CPU and memory overhead. So, it is hard to be made concurrent.
\item \emph{Small FIFO is too big}. The Small FIFO is designed to remove correlated
  references that typically occur within a short timeframe. However, 25\% of the cache is allocated to the Small FIFO, a design choice made 30 years ago when caches were much smaller. In today's storage systems, caches often exceed 10 gigabytes, and 25\% of it could hold more than half a million blocks. Even under
  highly concurrent workloads, with thousands of threads loading cold blocks into the cache, it is unlikely that such a large Small FIFO is necessary to filter out correlated references. Reducing the Small FIFO size would allow for a larger Main LRU, potentially improving the overall miss ratio.
\item \emph{Unnecessary miss for hot blocks}. To enter the Main LRU, a block must first be evicted from Small FIFO and placed in the Ghost FIFO. This causes extra misses for blocks that are actually hot. In a large cache with low miss ratio, this introduces a non-trivial miss ratio increase.
\end{itemize}

\subsection{Clock2Q}\label{subsec:clock2q}
\begin{figure}[t]
\centering
      \includegraphics[width=.775\linewidth]{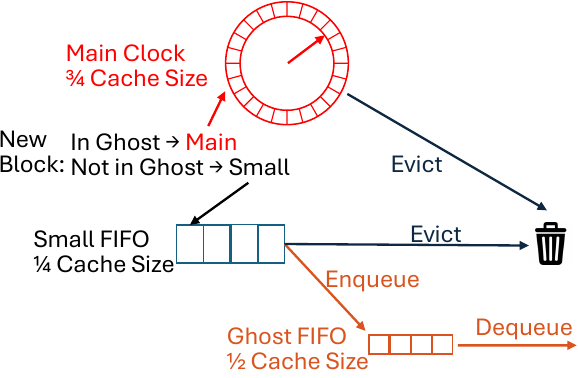}
    \caption{Illustration of Clock2Q replacement algorithm. \vspace{-0.8em}}\label{fig:clock2q}
\end{figure}

Clock2Q is the previous algorithm used in \productX. 
As shown in Figure~\ref{fig:clock2q}, the Clock2Q algorithm addresses the first limitation
of 2Q by replacing the Main LRU with a Main Clock ~\cite{clock2q-patent}. This modification lowers
the CPU and memory overhead and was used in \productX \verA, earlier versions of \productY and \productX
\verB ~\cite{clock2q-patent, clock2q-concurrent-patent}. Clock2Q still has the other two issues of 2Q (oversized Small FIFO and extra misses for hot blocks).

\subsection{S3-FIFO}\label{subsec:s3fifo}

\begin{figure}[t]
\centering
      \includegraphics[width=.775\linewidth]{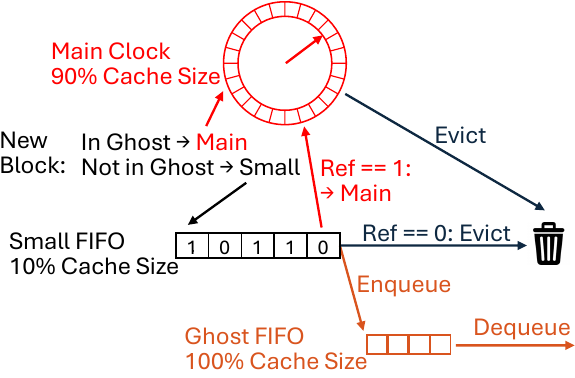}
    \caption{Illustration of S3-FIFO replacement algorithm. \vspace{-0.8em}}\label{fig:s3-fifo}
\end{figure}

S3-FIFO is the state-of-the-art algorithm~\cite{yang_fifo_2023-1}.
As shown in Figure~\ref{fig:s3-fifo}, it is structurally similar to Clock2Q, but the Small
FIFO now incorporates a Ref bit to track whether a block has been referenced while it
is in the Small FIFO. The size of each cache component is also adjusted: the Main
Clock size is increased from 75\% to 90\%, the Small FIFO size is reduced from 25\% to
10\%, and the Ghost FIFO size is increased from 50\% to 100\% of the cache size.

S3-FIFO uses the Ref bit in the Small FIFO to remove the extra miss caused by
hot blocks in 2Q and Clock2Q. When a block is accessed again after it enters
the Small FIFO queue, its Ref bit is set.
Later when this block is evicted from the Small FIFO queue, it
will be moved into the Main Clock directly because its Ref bit is set. The block can bypass the Ghost FIFO and be directly promoted to the Main Clock, eliminating the unnecessary eviction step in 2Q and Clock2Q.

S3-FIFO still uses the Ghost FIFO to identify blocks that are evicted from the
Small FIFO but later re-referenced.

S3-FIFO addresses all three issues of 2Q discussed in \S\ref{subsec:2q}. However,
it overcorrects the problem of the oversized Small FIFO by failing to handle correlated references altogether.
This is caused by the Ref bit in the Small FIFO. If a cold
block in the Small FIFO is accessed more than once due to correlated references,
its Ref bit will be set. As a result, the block is wrongly promoted to the Main Clock and displace other hot blocks. This is not a major issue for data caches, where correlated references are rare. However, it may not work well on metadata
cache, which has a lot more correlated references.

\subsection{Clock2Q+}\label{subsec:clock2qplus}

\begin{figure}[t]
\centering
      \includegraphics[width=.775\linewidth]{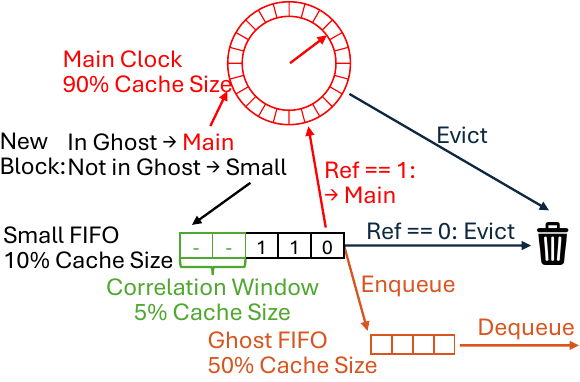}
    \caption{Illustration of Clock2Q+ replacement algorithm. \vspace{-0.8em}}\label{fig:clock2q-plus}
\end{figure}

As shown in Figure~\ref{fig:clock2q-plus}, Clock2Q+ is based on Clock2Q and
S3-FIFO, while addressing all the previously mentioned issues. The main difference between 
Clock2Q+ and S3-FIFO is the introduction of a ``Correlation Window'' in the Small FIFO.
The Correlation Window is set to 50\% of the Small FIFO size, or 5\% of the total cache size. 
When a block in the Small FIFO is accessed, if it is still within the Correlation Window (i.e., within the Correlation Window size of blocks from the head of the Small FIFO), its Ref bit will not be set. This ensures that if
correlated references occur within this short period, the block
will not be misclassified as hot. After the block moves beyond of the Correlation Window, the subsequent access sets the Ref bit, making sure that it will be directly promoted to the Main Clock when it reaches the end of the Small FIFO.

Additionally, Clock2Q+ reduces the Ghost FIFO size to 50\% of the total cache size, as empirical evaluations show that this performs as well as the previous 100\% allocation while reducing memory overhead.

\section{Making Clock2Q+ Production Ready}
\label{sec:production}
Before Clock2Q+, Clock2Q served as the core cache replacement algorithm used in \productX and \productZ for several years. 
%~\cite{vsan, vsanfs}
The improved Clock2Q+ algorithm has been implemented and will be included in future releases. This section discusses the engineering efforts required to make the cache replacement algorithm production-ready. 

\subsection{Implementation}\label{sec:impl}
To reduce memory fragmentation and enhance performance, Clock2Q+ eliminates the need for memory allocation after cache initialization. Each of the Main Clock, Small FIFO, and Ghost FIFO is represented as an array within a contiguous block of memory. Arrays of cache entries are allocated for the Main Clock and Small FIFO, and an array of cache keys is allocated for the Ghost FIFO.

To support the hash table for cache entry lookup, an array of pointers to cache entries is allocated as the hash buckets for the cache data, in both the Main Clock and Small FIFO. Similarly, an array is allocated for the hash buckets for entries in the Ghost FIFO. Each cache entry contains a next-pointer for to handle hash collisions.

\subsubsection{Concurrency}\label{ssec:concurrency}

A cache is often accessed by multiple threads concurrently, making efficient thread safety an essential design requirement. Because of the simple data structures in Clock2Q+ (chained hash tables, FIFO queues, and arrays), ensuring concurrency is straightforward.

A lightweight mutex lock is added to each hash bucket for the chained hash table. The lock
is held while operating on the linked list pointed by this bucket. Since
the load factor is low ($<1$), the average length of the linked list remains short. A
per-bucket lock provides a fine-grained protection. Our in-house performance tests have never shown
lock contention on these locks. 

For cache entries in the Main Clock and Cold FIFO, a lightweight mutex lock is assigned to each entry. The lock is held briefly while the entry is read or updated, as the operations are fast in-memory computations. When an
entry is evicted, and a disk I/O operation is issued to bring new content into this entry,
the entry is marked as ``Doing I/O'' and then inserted into the hash table. The
per-entry lock is then released during the whole duration of the I/O. If
another thread attempts to access the same block and finds it ``Doing I/O'', it will wait
on the entry until the I/O finishes.

Because each entry has a hash bucket lock and an entry lock, the lock ordering
must be consistent to avoid deadlock. However, different operations require
different lock orders:

\begin{itemize}
  \item \emph{Hash lock first} for cache lookup: A cache lookup must first lock
    the hash table to find the entry and then lock the cache entry.
  \item \emph{Entry lock first} for new entry insertion: When a new entry is
    added, the entry is first locked to update its status. Then this entry
    is inserted into the hash table, at which point the hash bucket must
    be locked. Similarly, when an entry is evicted, the entry is locked first, followed by locking the hash bucket for removal from
    the hash table.
\end{itemize}

Clock2Q+ follows the ``entry lock first'' order.
On every cache lookup,
after the hash bucket lock is obtained and the entry is found, the hash bucket
lock is released before the cache entry lock is taken. After the entry is locked, it is checked to ensure it has the expected key. Otherwise, it will
unlock the entry and retry. This flow is shown in
Figure~\ref{fig:lock-order}. In line 9, after the entry is found in the
hash table, this entry is evicted before it is locked in line 7. By comparing
the key of the entry in line 8, we can identify the race and retry. If the entry
cannot be found during the retry, it is treated as a cache miss.

\begin{figure}[ht]
  \begin{center}
    \begin{minipage}{.85\linewidth}
     \begin{lstlisting}[style=customc]
CacheGetEntry(key)
{
retry:
    LockHashBucket();
    entry = SearchHash(key);
    UnlockHashBucket();
    LockEntry(entry);
    if (entry.key != key) { // lost race
        UnlockEntry(entry);
        goto retry;
    }
    work on the entry;
    UnlockEntry(entry);
    ...
}
     \end{lstlisting}
    \end{minipage}
    \vspace{-1.6em}
         \caption{Lock Order Handling for Cache Lookup \vspace{-1.6em}}\label{fig:lock-order}
  \end{center}
\end{figure}

Another part of the concurrency design involves the head and tail in the Small FIFO and the
Ghost FIFO, and the clock hand in Main Clock. Since the Small and Ghost FIFO queues
are always full, a single index is used to indicate both their head and tail. All
these indices are atomic variables. Multiple threads can atomically increment
these variables so that they traverse through the FIFO or Clock array to find
eviction candidates or add new entries into the FIFO.

\subsubsection{Debugging}
The race condition presented in \S\ref{ssec:concurrency} is rare in real-world systems. However, it is important to test and verify its behavior. We developed a
race enforcement framework that can control any thread to pause on a ``breakpoint'' and resume when needed. We have a unit test using this framework that pauses one thread between line 6 and 7 in
Figure~\ref{fig:lock-order}, and let the second thread evict the cache entry the first thread is attempting to access. This forces the first thread to enter the race condition in lines 9 and 10, enabling us to test the handling of the race condition.

\subsubsection{Dirty Pages}\label{ssec:dirtypages}

The cache must handle dirty pages, which are often managed by Write-Ahead
Logs. Before log entries are flushed to persistent storage, the corresponding dirty
pages described by these log entries cannot be written
out~\cite{haerder1983principles}. Even if the dirty pages can be written out
and thus chosen as candidates for eviction, flushing them requires an I/O operation, making them much slower to evict compared to clean pages. Therefore,
Clock2Q+, as used in \productX and \productZ, skips dirty pages when selecting eviction candidates. This is equivalent to reinserting the dirty page at the head of the Main Clock or the Small FIFO. 

Another issue related to dirty pages arises when moving hot pages from the Small FIFO to
the Main Clock. If a dirty page at the end of the Small FIFO has its Ref bit set, it should be moved to the Main Clock. Although it is technically possible to move the page, the current implementation leaves it in the Small FIFO. Since the implementation uses an array instead of linked
lists for both the Small FIFO and the Main Clock, moving a page between them requires copying rather than simple pointer updates.
Furthermore, the transaction system tracks dirty pages using a dirty list for future flushing. Copying dirty pages is challenging because it requires locking the dirty list during the copy process.
The dirty page remains in the Small FIFO until it is revisited in a later cycle and is no longer dirty.
The impact of this change on the overall miss ratio is expected to be small. \S\ref{sec:dirty-sim-impact} evaluates and confirms that this change has a negligible effect on cache performance.

Since the Small FIFO is small, but the number of dirty pages can be large, there is a risk that, for some workloads, every page in the Small FIFO could be dirty. The default algorithm could enter an infinite loop, searching for an evictable page in the entire Small FIFO and failing. To address this, instead of scanning the entire Small FIFO, the algorithm scans only a fixed number of entries. If all scanned entries are dirty, it skips the remaining entries and checks the Main Clock for eviction candidates.

In our system, two flushing mechanisms are implemented to handle dirty pages:

\begin{enumerate}
    \item Time-based flushing: A time threshold is set. If dirty pages remain in the cache longer than this threshold, they are flushed.
    \item Proportion-based flushing: The cache maintains low and high watermark thresholds for the proportion of dirty pages. If the proportion of dirty pages in the cache exceeds the high watermark, dirty pages are flushed in order, from oldest to newest, until the proportion drops below the low watermark.
\end{enumerate}

\subsection{Live Cache Resizing}

A generic storage system like \productX must support a wide range of workloads. Some
workloads have large working set sizes and require a larger cache than the default
size. Dynamic cache growth and shrinking are therefore built into the system so that the
cache can handle requests without disruption while it is being grown or
shrunk.

The cache resizing mechanism must meet the following criteria:

\begin{itemize}
    \item Live: All regular cache operations must continue during resizing.
    \item Low CPU overhead: The additional CPU usage caused by resizing should not affect cache performance.
    \item Minimal data movement: The majority of the cache data should remain in place during resizing.
    \item Fine granularity: Cache size adjustments should be incremental, growing and shrinking by small amounts each time, avoiding large-scale reallocations.
    \item No memory fragmentation: Frequent cache resizing should not lead to fragmented system memory.
\end{itemize}

\subsubsection{Cache Growth}\label{sec:grow}

When the cache grows, all data structures in the cache must also grow, including arrays for the Small FIFO, Main Clock, and Ghost FIFO. During cache initialization, sufficient unused virtual address space is reserved at the end of each array. This allows the arrays to grow without requiring relocation. The first step in cache growth is to allocate additional space at the end of all these arrays.

After the newly grown entries are initialized, the cache size and hash bucket size are updated to reflect the new cache size. A background thread then begins traversing all cache entries, moving them from the old location to the new location. Each cache entry may reside in either the old or the new location during this process. Searching through two hash locations correctly is challenging because cache entries are not locked during lookups, and the background thread can move them at any point during the search. Our design shifts this complexity from lookup time to insertion time.

When a cache lookup occurs, it searches only the new hash location. If a false negative occurs because the entry still resides in the old hash location, the insertion routine will detect it, rehash it, and move it to the new location. Subsequently, the retry logic described in \S\ref{ssec:concurrency} will redo the lookup and succeed. This design simplifies the handling of race conditions: while cache entries are not locked during lookups, both cache entries and hash buckets are locked during insertion. As a result, locking both hash locations, searching, and moving entries becomes more manageable.

\subsubsection{Cache Shrinking}

Cache shrinking follows a similar approach to cache growth. The background thread first rehashes all cache entries into the new, smaller hash buckets and then frees the extra memory at the end of the hash bucket arrays. Data in the Ghost FIFO queue is easier to shrink because entries at the end of the array can be discarded.

Data at the end of the Main Clock and Small FIFO arrays must be handled differently. Clean pages can be discarded; however, dirty pages cannot be discarded. As discussed in \S\ref{ssec:dirtypages}, copying dirty pages in the cache is not straightforward. When a dirty page is encountered, the background thread triggers a transaction flush and retries later, with the hope that the page will be flushed and clean by the next check.

\subsection{Lessons Learned}

A cache replacement algorithm is a critical component of the data path in a storage system. Many storage products and modules at \companyX require effective cache replacement algorithms. The Clock2Q+ algorithm originated from a simple Clock2Q algorithm and evolved over the years, shaped by the needs of various products and requirements. It ultimately developed into Clock2Q+ and is now consolidated into a single source codebase, shared across multiple products.

In 2015, a new feature in \productX \verA required a CPU-efficient, scan-resistant cache replacement algorithm for the core data path running in the OS kernel. The expected cache hit ratio was close to 100\%, making the low CPU overhead of a Clock-based algorithm a suitable choice. The code operates as a state machine on a single core, so there was no need to handle locking or concurrency. Since the algorithm runs in kernel mode, debugging is challenging. Therefore, selecting a simple algorithm was a high priority. Additionally, the algorithm needed to be scan-resistant, as certain workloads periodically scan all blocks.

Although multiple scan-resistant Clock-based algorithms, such as Clock-Pro~\cite{clockpro-Jiang2005} and CAR/CART~\cite{car-Bansal2004}, are available, none of them was simple enough for consideration. The 2Q algorithm is one of the simplest algorithms that is also scan-resistant. However, it relies on LRU, which has a higher CPU overhead.

At this point, the Clock2Q design became straightforward: replace the LRU in 2Q with Clock, as Clock has been shown to behave similarly to LRU. Almost 10 years later, the S3-FIFO paper demonstrated that Clock performs better than LRU in many cases, which came as a pleasant surprise: Clock2Q may have a lower miss ratio and lower CPU and memory overhead than 2Q!

Another benefit of Clock, compared to Clock-Pro or CAR/CART, is that it does not require adjusting the size of the Clock or inserting/deleting elements in the middle of the Clock. As a result, arrays, instead of linked lists, can be used for the Main Clock, Small FIFO, and Ghost FIFO, which reduces memory overhead and further simplifies the implementation.

Around the same time, another new project, \productY, also required a cache algorithm. Since multiple instances of the cache were needed to store different types of data, a C++ template was used to generalize the cache implementation. Clock2Q was chosen again for its simplicity.

In this use case, concurrency was required because the cache was accessed by multiple threads. We started with a simple solution of using one giant lock to protect the entire cache. Although the lock was released when reading data from the disk during a cache miss, the giant lock still caused high lock contention. The next solution we tried was sharding. Sharding resolved the lock contention until the project's transaction system required reserving cache blocks so that each transaction had enough free cache space to finish without deadlocking the system. However, because blocks needed by a transaction could go to any shard, if one transaction needed $N$ blocks, it was only safe to reserve $N$ blocks in each shard, which was too wasteful to be practical.

We designed the highly concurrent version of Clock2Q discussed in \S\ref{ssec:concurrency} to replace sharding. After this change, cache access was no longer a performance bottleneck.

Since \productX supports a wide variety of workloads, its cache must support online resizing with minimal data movement. Because Clock2Q uses only arrays and a chained hash table, this simplicity enables us to support growth and shrinking with minimal changes.

After S3-FIFO was published, its similarity to Clock2Q became the most important factor that attracted our team to look deeper. While studying S3-FIFO, we began the work described in this paper with an initial hunch that adding a correlation window in the Small FIFO could combine the best aspects of Clock2Q and S3-FIFO and outperform both.

As the cache supports various workloads, a bug was discovered: when every page in the Small FIFO is dirty, the algorithm enters an infinite loop. This inspired us to study the impact of limiting the maximum number of cache entries the Cold FIFO head or clock hand needs to visit before it gives up, which is discussed in \S\ref{sec:clock-hand-upper-bound}.

When S3-FIFO was implemented in \productX, we found that dirty blocks needed to be moved from the Small FIFO to the Main Clock, which was never necessary in Clock2Q. We chose simplicity over accuracy, as discussed in \S\ref{ssec:dirtypages}.

The above evolution of Clock2Q+ illustrates a typical process of how an academic algorithm is adapted to real systems. Minimal engineering effort is applied at each stage to avoid premature optimization. An algorithm with balanced characteristics (simplicity, reasonable hit ratio, low CPU overhead, high concurrency, etc.) is desired. Among all these traits, simplicity is of paramount importance in production systems, although it is rarely discussed in academia. A cache replacement algorithm is a type of algorithm where simplicity is especially critical. If there is a bug in the cache replacement algorithm, it is hard to determine whether it is a bug or the algorithm simply does not work well under that workload. It is even harder to reproduce and debug in production systems. This is because real workloads are not easily accessible, and it is difficult to set up and run repeatable workloads with realistic access patterns. Commonly available benchmarks use simple distributions (such as uniform random or Zipf), which do not represent realistic access patterns. In our case, the initial choice of a simple design proved to be highly beneficial, especially when cache resizing was implemented and tested many years later.

\section{Evaluation}\label{sec:eval}

In this section, we discuss Clock2Q+ in relation to all five goals listed in \S\ref{sec:intro} and focus on evaluating its miss ratio against other algorithms.

Clock2Q+ has the same low CPU overhead as any clock-based algorithm on a cache hit, requiring only a hash lookup and a Ref bit update. It also has high multi-core scalability because every cache entry and hash bucket has a separate lock. In any performance testing in \productX \verB and \productY, cache lock contention never appeared as a bottleneck. The implementation of Clock2Q+ is straightforward because it only uses a few arrays (for the Clock, FIFOs, and hash buckets) and singly linked lists (for hash table collisions). This makes it simpler than any other scan-resistant algorithm. Additionally, Clock2Q+ requires no manual tuning, as its default parameter is robust (see \S\ref{sec:eval:sensitivity}).

In this section, we compare Clock2Q+ to state-of-the-art cache replacement algorithms on different workloads, including metadata, data, and non-block workloads. We study the impact of dirty blocks and evaluate whether the simplifications in \S\ref{ssec:dirtypages} are effective. We further analyze why it achieves higher cache efficiency. Additionally, we study the sensitivity of its parameters to ensure its robustness.

\subsection{Experiment Setup}
\label{sec:eval:set}
\noindent\textbf{Workloads.} Since \productX is deployed on enterprise customers' hardware, we cannot conduct experiments using customers' data due to privacy and compliance reasons. Therefore, we used open-source block cache datasets from CloudPhysics~\cite{waldspurger_efficient_2015}. The CloudPhysics dataset contains 106 traces spanning 7 days, with 2,114 million requests for 82 TB of data. Since Clock2Q+ is designed for metadata caching in \productX, and none of the open-source traces were collected from metadata caches, we derived metadata traces from the production data traces using the method described in \S\ref{sec:cache-trace-type:convert}: dividing the LBNs in the data traces by 200. The fan-out of 200 was chosen to match the fan-out of the real \btree in \productX \verB.

\noindent\textbf{Benchmarks.} We implemented Clock2Q+ in the cache simulator libCacheSim~\cite{libCacheSim}. This ensures a fair and accurate comparison with state-of-the-art algorithms. We evaluated different cache replacement algorithms at four different cache sizes: 0.005, 0.01, 0.05, and 0.1 of the trace footprint.

\noindent\textbf{Baseline and Metrics.} 
We compared Clock2Q+ with state-of-the-art replacement algorithms. Since previous work has shown that Clock achieves both lower overhead and a lower miss ratio than LRU~\cite{yang_fifo_2023}, we used Clock as the baseline and calculated the miss ratio improvement as 
\begin{equation}
\text{Improvement}_{\text{algo}}= \left( \frac{\text{MR}_{\text{Clock}} - \text{MR}_{\text{algo}}}{\text{MR}_{\text{Clock}}} \right)
\end{equation}

\subsection{Metadata Trace Fidelity}

\begin{figure}[ht]
  \begin{center}
    \includegraphics[width=.64\columnwidth]{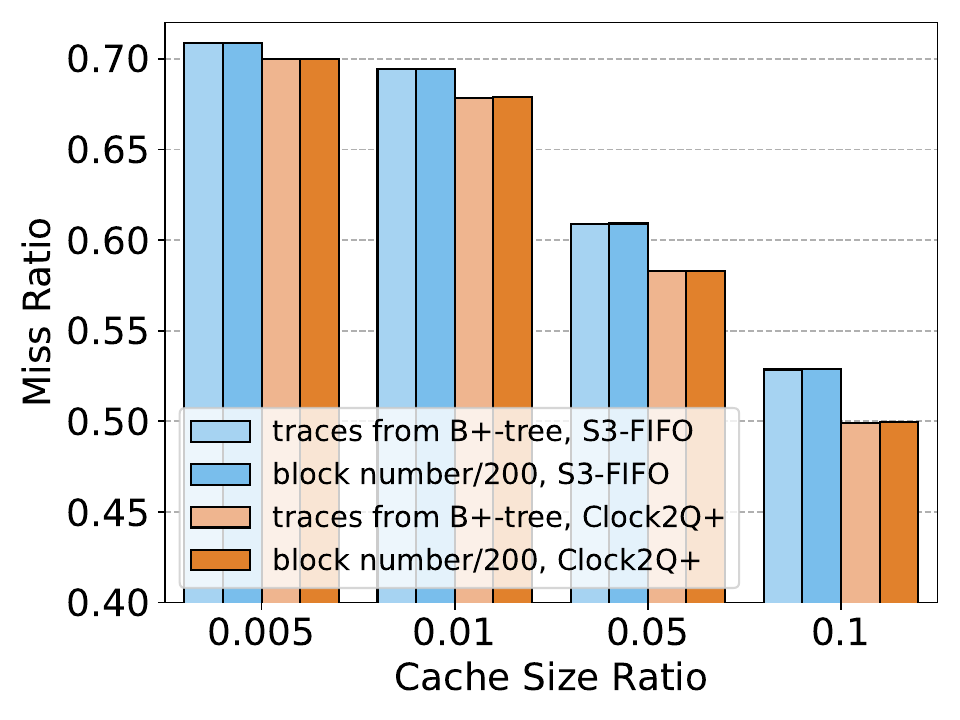}
    \caption{Clock2Q+ and S3-FIFO show almost identical miss ratios on derived metadata traces and traces collected from TLX \btree leaf block accesses.\vspace{-0.8em}}
    \label{fig:tlx-leaf-pattern}
  \end{center}
\end{figure}

We validate the realism of our metadata trace produced by our simple derivation method by generating and comparing two metadata traces. 
In the first trace, we created a \btree using the TLX library~\cite{TLX}, replayed a data trace, and recorded the accessed leaf blocks. 
In the second trace, we applied the method described above: for each request in the data trace, we divided the LBN by 200. 
We evaluated S3-FIFO and Clock2Q+ on the two traces. 
Figure~\ref{fig:tlx-leaf-pattern} shows that all differences are less than 0.01\% between the two traces. This indicates that the proposed method of dividing LBN by fan-out is an effective approach for deriving metadata traces from data traces.

\subsection{Miss Ratio Improvements}

\begin{figure*}[t]
    \centering
    \begin{subfigure}[t]{0.48\textwidth}
    \includegraphics[width=\columnwidth]{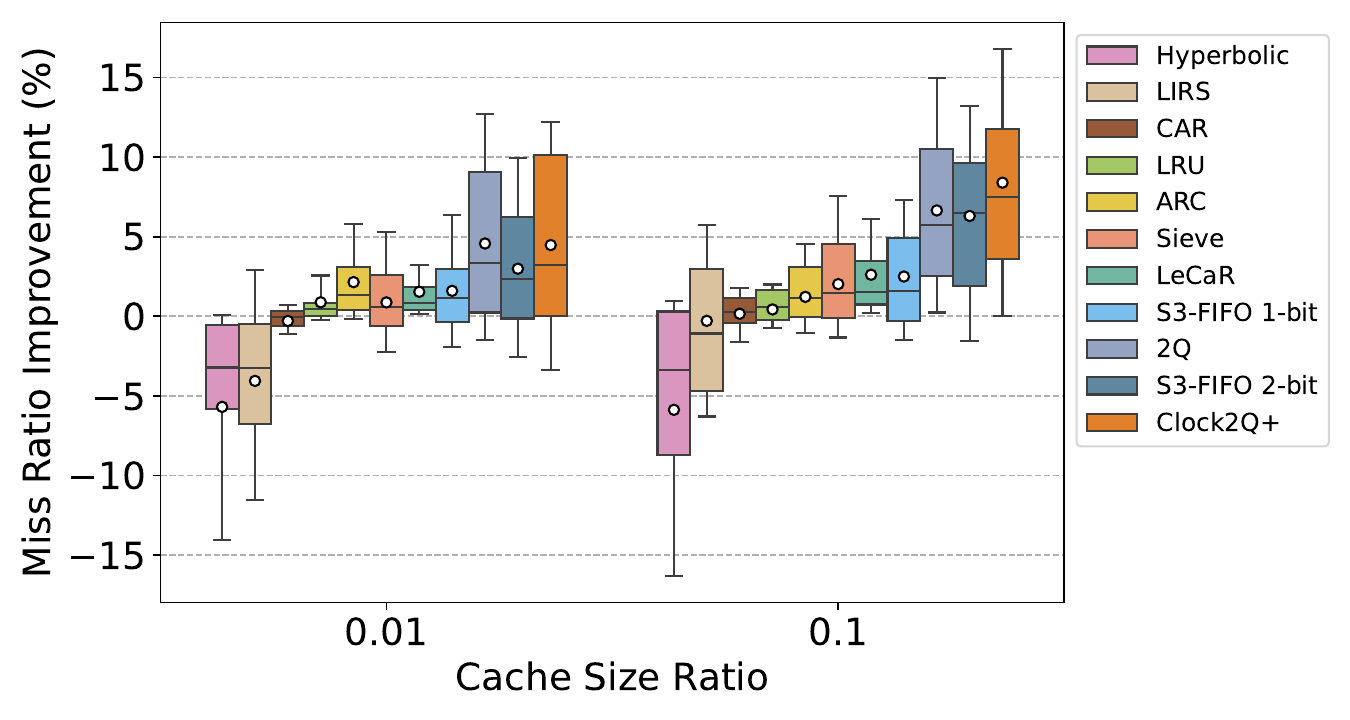}
    \caption{Metadata traces}
    \label{fig:all_alg_meta}
    \end{subfigure}
    \begin{subfigure}[t]{0.48\textwidth}
    \includegraphics[width=\columnwidth]{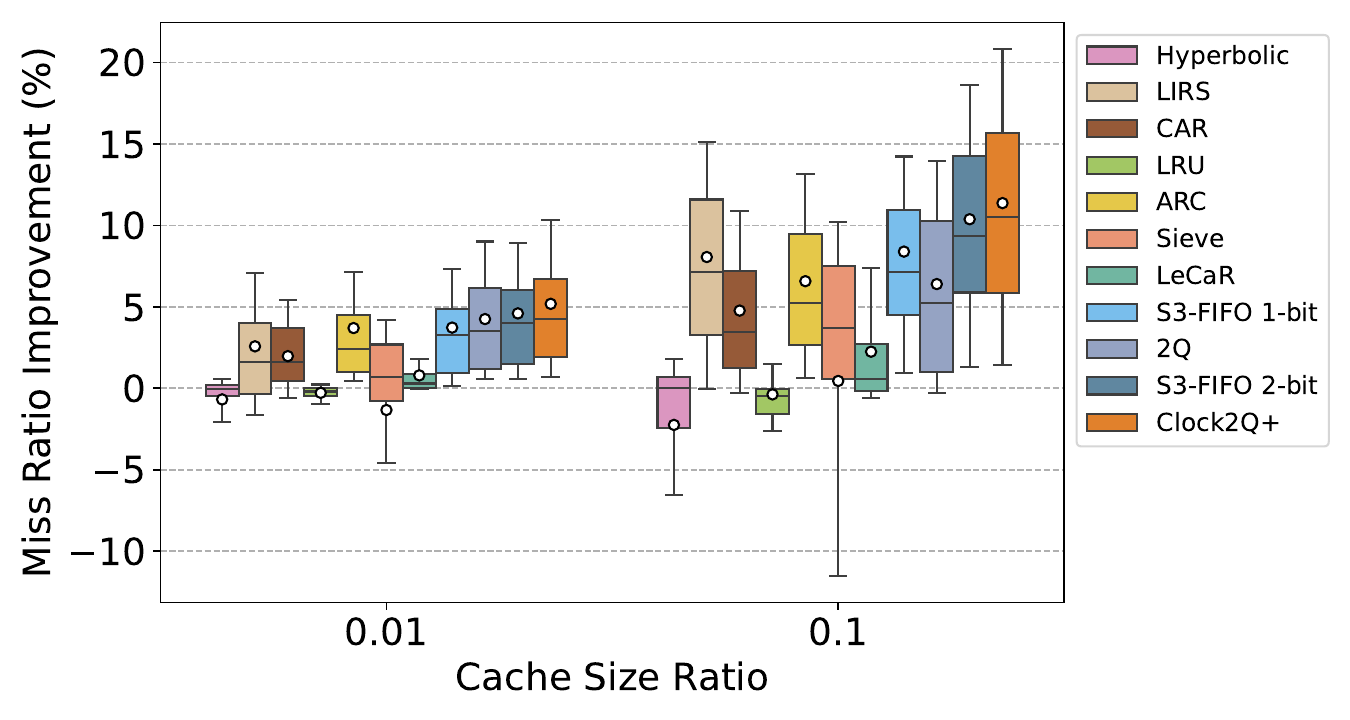}
    \caption{Data traces}
    \label{fig:all_alg_data}
    \end{subfigure}
    \caption{Comparing Clock2Q+ with state-of-the-art eviction algorithms. \vspace{-0.8em}}\label{fig}
\end{figure*}

\noindent\textbf{Metadata traces.} Figure~\ref{fig:all_alg_meta} compares Clock2Q+ with ten state-of-the-art cache replacement algorithms on the metadata traces. 
We observe that most algorithms do not significantly outperform Clock; that is, they do not show positive improvements in the figure. 
This is because these algorithms struggle to handle correlated references effectively. As a result, blocks with correlated references are often misclassified as hot, leading to suboptimal eviction decisions in these algorithms.

We find that only S3-FIFO 2-bit (default S3-FIFO), 2Q, and Clock2Q+ demonstrate significant improvements across the two cache sizes. 
S3-FIFO 2-bit uses a 2-bit frequency counter. Blocks in S3-FIFO 2-bit need to be re-referenced at least twice before they can be moved to the Main Clock, which effectively reduces the likelihood of cold blocks with correlated references entering the Main Clock. In comparison, S3-FIFO 1-bit admits blocks with only a single re-reference from the Small FIFO to the Main Clock. As a result, it is more susceptible to correlated references and exhibits a higher miss ratio compared to S3-FIFO 2-bit. 
In 2Q, blocks in the Small FIFO are always evicted without being reinserted into the Main Clock. This approach also addresses the correlated reference problem, as the frequency of block access does not influence this behavior. 

Clock2Q+, as described in \S\ref{subsec:clock2qplus}, is specifically designed to handle workloads with correlated references, utilizing the correlation window filter to prevent cold blocks from being moved into the Main Clock. 
Among all algorithms, Clock2Q+ achieves the best miss ratio improvement across all tested cache sizes. For the larger cache size, S3-FIFO 2-bit performs second-best, and Clock2Q+ outperforms it by up to 28.5\% in miss ratio under this setup.

\noindent\textbf{Data traces.} For comparison, we evaluated the same set of algorithms on data traces. 
Figure~\ref{fig:all_alg_data} shows that most of the state-of-the-art algorithms provide substantial miss ratio improvements compared to the baseline (Clock). This differs from the previous observation in Figure~\ref{fig:all_alg_meta} but aligns with expectations, as these algorithms are designed for data cache workloads. While the level of improvement varies across different replacement algorithms, Clock2Q+ still demonstrates the highest improvement among all. Despite being designed for metadata cache workloads, Clock2Q+ also performs exceptionally well on data cache workloads. Furthermore, this suggests that some data cache workloads may also exhibit correlated references, highlighting the broader applicability of Clock2Q+.

\begin{figure}[ht]
  \begin{center}
    \includegraphics[width=\columnwidth]{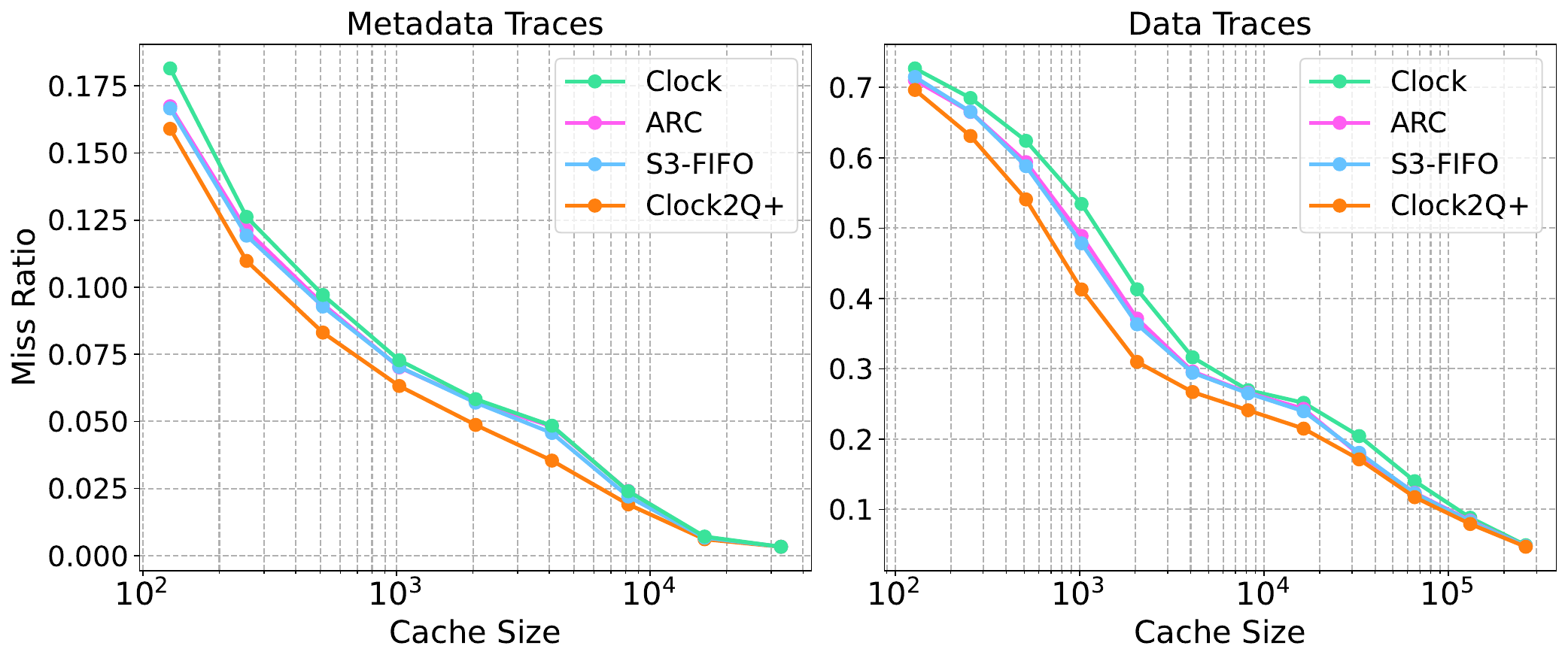}
    \caption{Miss-ratio curves for CloudPhysics w56. The x-axis is cache size (log scale, unit is byte), and the y-axis is miss ratio. Lower curves indicate better performance (fewer misses) at a given cache size.}
    \label{fig:mr_curve}
  \end{center}
\end{figure}

\noindent\textbf{Consistent improvement across cache sizes.}
Figure~\ref{fig:mr_curve} summarizes the miss–ratio curves for the CloudPhysics w56 workload. The left shows metadata and the right row shows data. Within each subplot, the four curves compare Clock, ARC, S3-FIFO, and Clock2Q+ on the same workload and object set.

Across all four panels, Clock2Q+ consistently has the best curve: it lies strictly below the existing algorithms over on the \textbf{entire} cache–size range (from 1\% to 100\%).
On metadata traces, Clock2Q+ provides a visible downward shift in the curve. For data blocks, the absolute miss ratios are higher. Here Clock2Q+ again has the lowest miss ratios over entire range of cache sizes.

In summary, Clock2Q+ consistently achieves lower miss ratios than state-of-the-art algorithms, with the improvement being more pronounced in the metadata traces.

\subsection{The Secret Behind High Efficiency}
To understand the high efficiency of Clock2Q+, we explore the differences in blocks inserted into the Main Clock and the Ghost FIFO in S3-FIFO and Clock2Q+. We use one trace as a case study throughout this section. 
We plotted the Probability Density Function (PDF) of the next reuse distance (NRD) of blocks moved to the Main Clock and to the Ghost FIFO from the Small FIFO. 
A small NRD indicates that the block will be accessed soon after leaving the Small FIFO and should ideally stay in the Main Clock to reduce cache misses. 
A large NRD suggests that the block will not be reused soon and can be evicted. 

\begin{figure}[ht]
  \begin{center}
    \begin{subfigure}[t]{0.49\columnwidth}
      \centering
      \includegraphics[width=\textwidth]{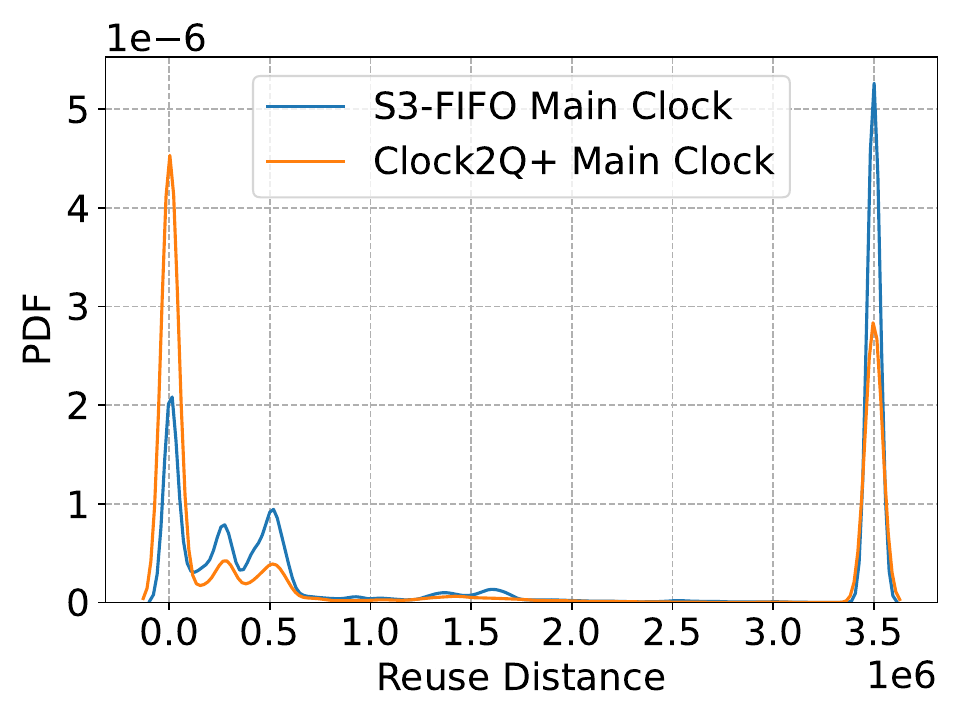}
      \caption{Blocks to Main Clock}
      \label{fig:main_pdf}
    \end{subfigure}
    \begin{subfigure}[t]{0.49\columnwidth}
      \centering
      \includegraphics[width=\textwidth]{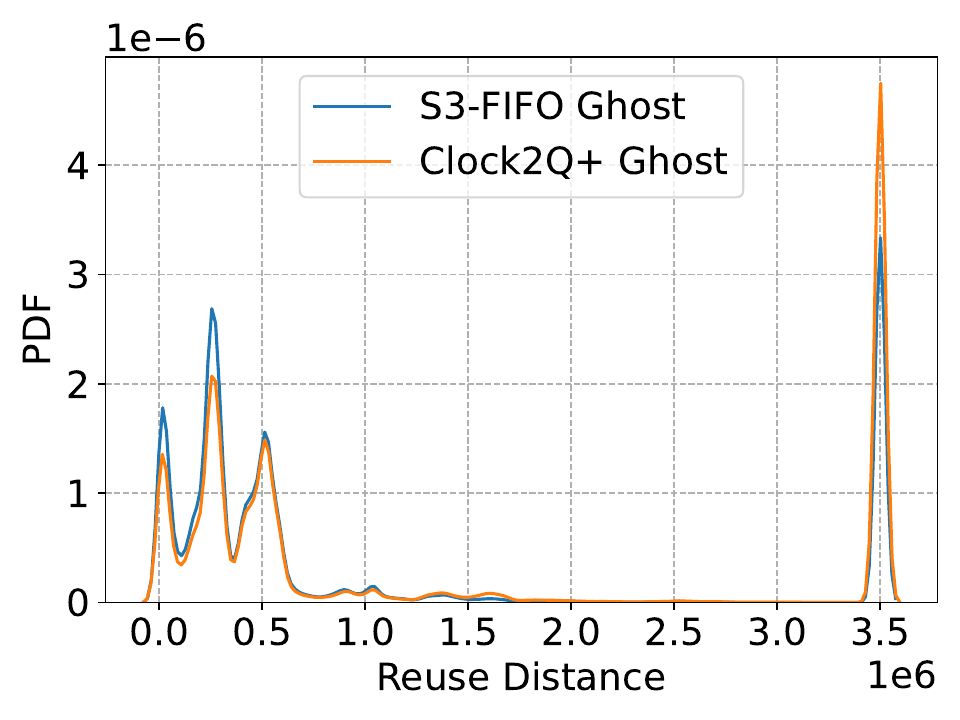}
      \caption{Blocks to Ghost FIFO}
      \label{fig:ghost_pdf}
    \end{subfigure}
    \caption{PDF of Next Reuse Distance (NRD) of blocks from Small FIFO. Clock2Q+ moves more cold blocks (large NRDs) to Ghost FIFO and moves more hot blocks (small NRDs) to the Main Clock. The rightmost peak represents blocks that are not accessed again.\vspace{-0.8em}}
  \label{fig:combined_pdf}
  \end{center}
\end{figure}

Figure~\ref{fig:main_pdf} shows that, compared to S3-FIFO, Clock2Q+ moves a larger portion of blocks with \textit{small} NRD to the Main Clock, ensuring these hot blocks remain in the cache longer. Figure~\ref{fig:ghost_pdf} shows that fewer such blocks end up in the Ghost FIFO, reducing the likelihood of mistakenly evicting blocks that will be reused soon. The results demonstrate that Clock2Q+ makes better decisions by keeping hot blocks in the Main Clock and evicting cold blocks to the Ghost FIFO.

Table~\ref{tab:block_counts} shows the number of blocks that Clock2Q+ and S3-FIFO moved into the Main Clock and the Ghost FIFO from the Small FIFO. 
Compared to S3-FIFO, Clock2Q+ moves significantly fewer blocks (less than $\frac{1}{4}$) from the Small FIFO into the Main Clock. 
At the same time, the number of blocks moved from the Cold FIFO to the Ghost FIFO is about the same between the two algorithms.
These findings highlight the importance of selecting the correct blocks to move from the Small FIFO to the Main Clock, as this decision will impact the miss ratio. By being more selective about what enters the Main Clock, Clock2Q+ reduces ``pollution'' in the Main Clock.
This suggests that the correlation window of Clock2Q+ is an effective mechanism for deciding which blocks should be promoted to the Main Clock when evicting from the Small FIFO.

\begin{table}[ht]
    \small
    \centering 
    \caption{Counts of inserted block in the Main Clock and Ghost FIFO for Clock2Q+ and S3-FIFO.}
    \label{tab:block_counts}
    \begin{tabular}{l|r|r|r}
    \hline
     & \textbf{Small $\shortrightarrow$} & \textbf{Small $\shortrightarrow$ } & \textbf{Ghost $\shortrightarrow$}\\
    \textbf{Algorithm} & \textbf{Main} & \textbf{Ghost} & \textbf{ Main}\\
    \hline
    Clock2Q+ & 20,762 & 186,959 & 33,068\\
    S3-FIFO  & 88,529 & 126,897 & 29,909\\
    \hline
    \end{tabular}    
\end{table}

\subsection{Impact of Production Optimizations}\label{sec:eval:production}
\subsubsection{Dirty Blocks}\label{sec:dirty-sim-impact}

In Clock2Q+, when the Small FIFO searches for an eviction candidate, it skips dirty blocks. After a threshold of dirty blocks is skipped, it gives up searching in Small FIFO and directly inserts the new block into the Main Clock. This avoids a dead loop when all blocks in the Small FIFO are dirty.

As discussed in \S\ref{ssec:dirtypages}, when a dirty block is evicted from the Small FIFO, if its Ref bit is set, it should be moved to the Main Clock. However, implementing such movement in \productX is not trivial. Therefore, the dirty block is re-inserted into the Small FIFO and will be checked again when it becomes the eviction candidate next time. Whether and by how much such optimizations for handling dirty blocks impact the miss ratio is investigated in this section.

We set the dirty block flush time to 30 seconds. The low and high watermarks for flushing dirty blocks are set to 10\% and 20\% of cache size, respectively, which are the same as the values\footnote{\texttt{dirty\_background\_ratio} and \texttt{dirty\_ratio} in \texttt{/proc/sys/vm/}, respectively} used in the Linux kernel page cache.

\begin{figure}[ht]
  \begin{center}
    \includegraphics[width=0.7\columnwidth]{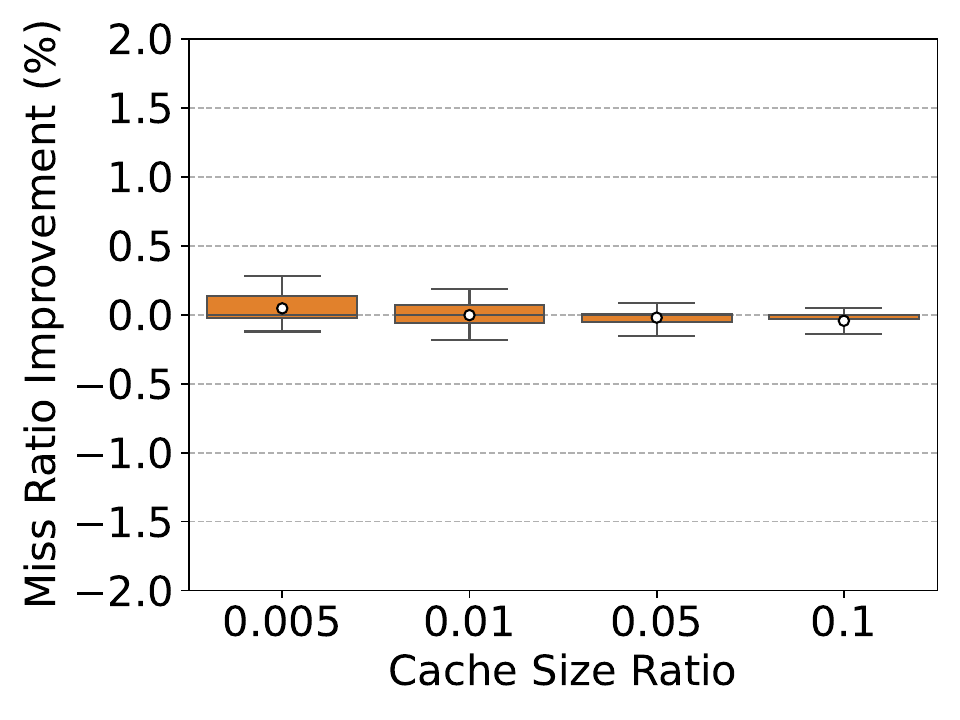}
    \caption{Impact of simplified dirty block handling. Miss ratio improvement is calculated against Clock2Q+ that moves dirty pages from the Small FIFO to the Main Clock. \vspace{-0.8em}}\label{fig:skip-dirty-impact}
  \end{center}
\end{figure}

Figure~\ref{fig:skip-dirty-impact} shows the impact of simplified dirty block handling on 106 CloudPhysics traces. 
A negative value indicates that the miss ratio is increased for the trace. 
We observe that the simplified dirty block handling has minimal impact on most traces. 
At the smallest cache size, we observe that skipping dirty blocks can reduce the miss ratio for some traces. We conjecture that when the cache size is very small, the Small FIFO is not large enough, causing some potentially hot blocks to be evicted from the Small FIFO. Therefore, skipping dirty blocks in the Small FIFO helps avoid evicting those hot blocks. At larger cache sizes, we observe that simplified dirty block handling slightly increases the miss ratio for some traces. However, overall, the impact is negligible. 
This observation justifies the simplification of dirty block handling in real systems.

\subsubsection{Limiting Clock Hand Movement}\label{sec:clock-hand-upper-bound}
\begin{figure}[ht]
    \centering
    \begin{subfigure}[t]{.49\linewidth}
      \includegraphics[width=\columnwidth]{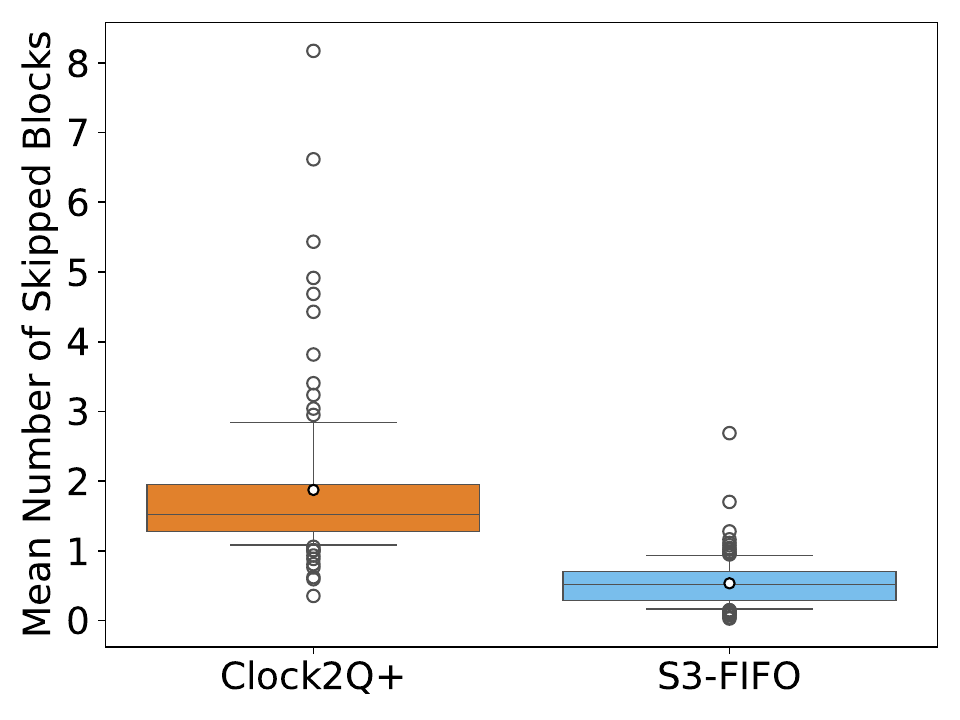}
      \caption{Mean skipped block count distribution}
      \label{fig:clock_hand_stats}      
    \end{subfigure}
    \begin{subfigure}[t]{.49\linewidth}
    \includegraphics[width=\columnwidth]{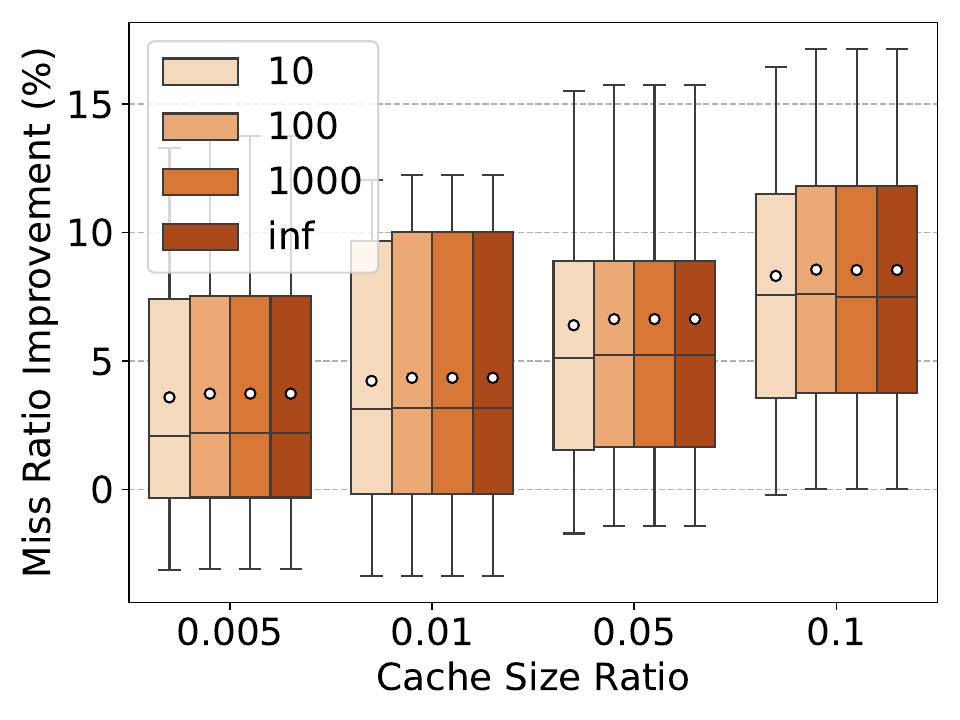}
    \caption{Impact of limiting the number of skipped blocks per eviction}
    \label{fig:clock_hand_upperbound}
    \end{subfigure}
    \caption{Impact of reinsertion limit on miss ratio.\vspace{-0.8em}}
    \label{fig:eval:reinsertion}
\end{figure}

When evicting from the Main Clock, a block can be skipped if its Ref bit is set. The number of skipped blocks can be high during a single eviction, which can potentially lead to increased CPU usage. 
Figure~\ref{fig:clock_hand_stats} shows the mean number of skipped blocks across the 106 CloudPhysics traces. We observe that it is low (<2) for most traces for both Clock2Q+ and S3-FIFO. Compared to S3-FIFO, Clock2Q+ has more skipped blocks per eviction. This is likely because Clock2Q+ keeps more hot blocks in the Main Clock, which results in more skipped blocks. 

To limit the number of skipped blocks, we set a threshold per eviction at 10, 100, 1000, and $\infty$ (no restriction). Once the reinsertion limit is reached, a block is forcibly evicted from the Main Clock if it is not dirty, regardless of its Ref bit value. 
Figure~\ref{fig:clock_hand_upperbound} shows the differences between the various thresholds. We observe that limiting the maximum number of skipped blocks to 10 only slightly increases the miss ratio. This suggests that limiting the number of skipped blocks to a small value, e.g., 10, is a safe choice for production systems.

\subsection{Sensitivity and Versatility Analysis}\label{sec:eval:sensitivity}

\begin{figure}[ht]
\centering
\begin{minipage}[t]{.49\linewidth}
    \includegraphics[width=\linewidth]{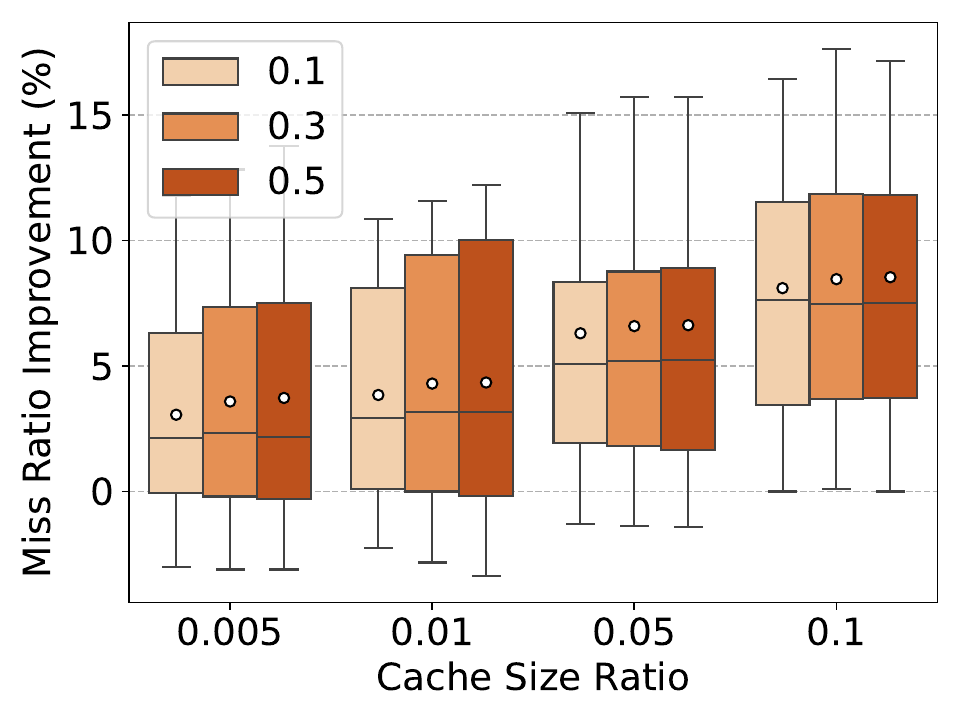}
    \caption{Impact of\\correlation window size.\vspace{-0.8em}}\label{fig:sensitivity}    
\end{minipage}
\begin{minipage}[t]{.49\linewidth}
    \includegraphics[width=\linewidth]{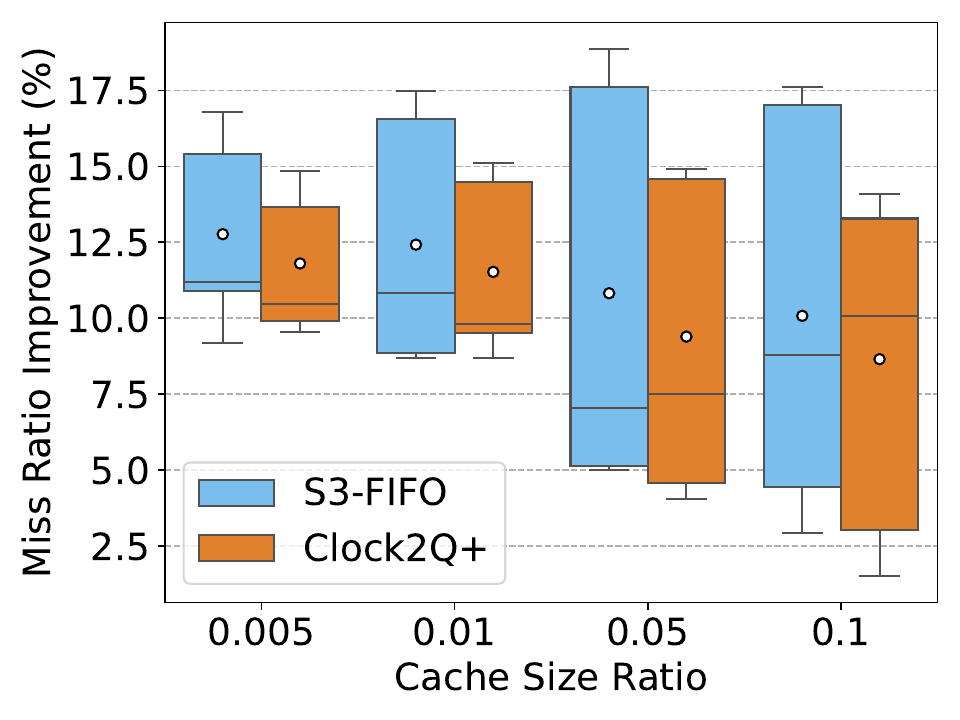}
    \caption{Miss ratio improvements on non-block traces.\vspace{-0.8em}}\label{fig:other_traces_data}
\end{minipage}
\end{figure}

Clock2Q+ uses half of the Small FIFO as the correlation window size. 
In this section, we evaluate the impact of the correlation window size. 
Figure~\ref{fig:sensitivity} shows the results for three different correlation window sizes (10\%, 30\%, and 50\% of the Small FIFO). 
At small cache sizes, a larger correlation window slightly improves the miss ratio; however, the improvement is limited at larger cache sizes. Our conjecture is that the correlation window size is an inherent property of the workload and should not vary significantly with cache sizes. Therefore, at small cache sizes, a larger correlation window is beneficial. 
Overall, the results indicate that the miss ratio is not very sensitive to the correlation window size, as all configurations provide significant miss ratio improvements over the baseline.

So far, we have only evaluated block traces and have shown that Clock2Q+ achieves a lower miss ratio than state-of-the-art algorithms on both derived metadata traces and the original data traces. We also explore the efficiency of Clock2Q+ on non-block cache workloads. We used three open-source cache datasets~\cite{yang_fifo_2023-1}, namely, the Wikimedia, Meta Key-Value, and Tencent Photo datasets. In total, there are 10 traces. 
Figure~\ref{fig:other_traces_data} shows that the miss ratio improvement of Clock2Q+ is slightly smaller than that of S3-FIFO on this dataset of key-value and object traces. We conjecture that this is because these workloads have limited correlated references, and the correlation window in Clock2Q+ does not provide enough benefit. Instead, it causes some potentially hot objects to be evicted from the Small FIFO.

\section{Related Work}\label{sec:bg}

%-------------------------------------------------------------------------------

\subsection{Cache Replacement Algorithms}

Cache replacement algorithm is a research area that spans decades with numerous literature~\cite{megiddo_arc_2003, zhang_sieve_2024, jiang_lirs_2002, tang_ripq_2015, johnson_2q_1994, carr_wsclock_1981, yang_gl-cache_2023, vietri_driving_2018, einziger_tinylfu_2017}. 
LRU is the most well-known and widely used algorithm because it is easy to understand and implement~\cite{lru-Effelsberg1984}. Clock approximates LRU with a lower overhead and a lower miss ratio~\cite{clock-Corbato1969, yang_fifo_2023-1}. Both LRU and Clock are not scan-resistant.

Many advanced algorithms were designed to make the cache scan-resistant. LIRS~\cite{jiang_lirs_2002} uses the distance between the last two references of a block to determine whether a block is hot. 
LIRS2~\cite{lirs2-21} improves this by using the sum of the distance to a block's last two recent accesses. 
Clock-Pro~\cite{clockpro-Jiang2005} is a Clock variation of LIRS where the CPU overhead of cache hit is small by setting a Ref bit during cache hit. 
DLIRS~\cite{dlirs-li18} dynamically tunes LIRS' parameters to make it work well in different workloads. 
ARC~\cite{megiddo_arc_2003} uses two queues with adaptive size to store blocks with recency and frequency. CAR and CART are the Clock variations of ARC~\cite{car-Bansal2004}. 
LRU-K~\cite{lruk-O1993} uses the last $kth$ access to a block to determine a block's hotness. When K is 1, it becomes LRU. 
SEQ~\cite{seq-glass97} detects long sequential accesses and handles them differently. 
SLRU~\cite{slru-94, huang_analysis_2013} split the LRU list into two parts: one holds blocks accessed only once, and one holds hot blocks. 
EELRU~\cite{eelru-99} keeps track of block numbers of blocks that are just evicted out of the cache to detect sequential scan or large loop patterns and evict blocks earlier to improve the hit ratio. 
LFU~\cite{lfu-Sleator1985} uses the frequency to determine the hot blocks that should stay in the cache. 
It tends to keep once frequently accessed blocks in the cache, even after they become cold later. 
Variations of LFU such as LFU$^*$~\cite{arlitt-thesis-96}, LFU-Aging~\cite{robinson90}, and LFU$^*$-Aging~\cite{willick-eager-bunt-93} are proposed to overcome this problem. 
FBR~\cite{robinson90} is similar to LFU but does not count correlated references. 
LRFU~\cite{lrfu-Lee1999} combines recency and frequency into one parameter and presents a spectrum of algorithms in between LRU and LFU. 
LeCaR~\cite{vietri_driving_2018} uses machine learning to determine the weight between LRU and LFU based on workload. However, LeCaR is not scan-resistant.
Cacheus~\cite{rodriguez_learning_2021} solves this by learning between two variations of LRU and LFU (SR-LRU and CR-LFU). 
GDSF~\cite{cherkasova_improving_1998} considers data size together with frequency for eviction.
TinyLFU uses a Bloom Filter and the Count Min Sketch to count the number of references to blocks before admitting them into the main cache~\cite{einziger_tinylfu_2017}. 
The \emph{Mr. cache} algorithm can be applied to any algorithm by splitting a cache into multiple regions to reduce the CPU contention~\cite{wenguang-thesis}.
LRB~\cite{song_learning_2020}, HALP~\cite{song_halp_2023}, and 3L-cache~\cite{3lcache} use machine learning to rank objects at eviction. 
GL-Cache~\cite{yang_gl-cache_2023} uses machine learning to rank object groups for group-level evictions. 
LHD~\cite{beckmann_lhd_2018} uses probability to decide eviction candidates. 
Hyperbolic~\cite{blankstein_hyperbolic_2017} uses frequency over age as a metric to rank eviction candidates. 
S3-FIFO~\cite{yang_fifo_2023-1} uses a small FIFO queue to quickly filter out unpopular objects before they enter the main Clock queue. 
SIEVE~\cite{zhang_sieve_2024} is a new Clock-like algorithm that uses a different hand movement to achieve ``quick demotion'' and ``lazy promotion''. 
Some algorithms use application hints to make replacement decisions, such as DEAR~\cite{choi-dear-98}, AFC~\cite{choi-afc-00}, UBM~\cite{kim-ubm-00}, ILRU and OLRU~\cite{sacco87}, Hot Set~\cite{sacco86}, and QLSM~\cite{chou-85}. 

\subsection{Correlated References}\label{sec:cor-ref}

Correlated references were identified in the work of FBR~\cite{robinson90}. Later, multiple algorithms were designed to identify and filter out correlated references. 2Q~\cite{johnson_2q_1994} uses the Small FIFO to filter them out. However, it evicts all the blocks exiting the FIFO queue, causing one cache miss for every hot block in the main Clock. 
CART~\cite{car-Bansal2004} is a variation of CAR but can handle correlated references. Its idea is similar to 2Q because it uses the length of $T_1$ (similar to the small FIFO of 2Q) as the correlation window size. If a block is found on one of its ghost queue $B_1$, the block is marked as outside of the correlation window. This is similar to 2Q where an extra miss is generated for hot blocks.
Clock2Q+ uses a better approach than CART and 2Q by not generating extra misses when identifying blocks outside of the correlation window.

\section{Conclusion}\label{sec:conclusion}

Metadata cache workloads often exhibit correlated references. We present a simple and effective cache replacement algorithm, Clock2Q+, designed for metadata caches and implemented in multiple production systems at \companyX. Clock2Q+ achieves a lower miss ratio on metadata traces than state-of-the-art cache replacement algorithms and provides up to a 28.5\% lower miss ratio than S3-FIFO. Besides a lower miss ratio, Clock2Q+ also has low CPU and memory overhead, is highly concurrent, requires no tuning, and is easy to implement, which is one of the most important characteristics desired in a production system.

\bibliography{paper1,paper2,paper3}{}

@Article{sacco86,
  author =       {Giovanni Maria Sacco and Mario Schkolnick},
  title =        {Buffer management in relational database systems},
  journal =      {ACM Transactions on Database Systems (TODS)},
  year =         {1986},
  OPTkey =       {},
  OPTvolume =    {11},
  OPTnumber =    {4},
  OPTpages =     {473-498},
  OPTmonth =     {December},
  OPTnote =      {},
  OPTannote =    {}
}

@InProceedings{chou-85,
  author =       {Hong-Tai Chou and David J. DeWitt},
  title =        {An evaluation of buffer management strategies for relational database systems},
  booktitle = {Proceedings of the 11th International Conference on Very Large Data Bases (VLDB 1985)},
  year =      1985,
  pages =     {174-188},
  month =     {August},
  address =   {Stockholm, Sweden}}

@InProceedings{sacco87,
  author =       {Giovanni Maria Sacco},
  title =        {Index access with a finite buffer},
  booktitle = {Proceedings of the 13th International Conference on Very Large Data Bases (VLDB 1987)},
  year =      1987,
  pages =     {301-309},
  month =     {September},
  address =   {Brighton, England}}

@InProceedings{kim-ubm-00,
  author =       {Jong Min Kim and Jongmoo Choi and Jesung Kim and Sam H. Noh and Sang Lyul Min and Yookun Cho and and Chong Sang Kim},
  title =        {Unified buffer management scheme that exploits sequential and looping references},
  booktitle = {Proceedings of the 4th USENIX Symposium on Operating Systems Design and Implementation (OSDI 2000)},
  year =      2000,
  month =     {October},
  address =   {San Diego, CA}}

@InProceedings{choi-afc-00,
  author =       {Jongmoo Choi and Sam H. Noh and Sang Lyul Min and and Yookun Cho},
  title =        {Towards application/file-level characterization of block references: A case for fine-grained buffer management},
  booktitle = {Proceedings of the 2000 ACM SIGMETRICS International Conference on Measurement and Modeling of Computer Systems (SIGMETRICS 2000)},
  year =      2000,
  pages =     {286-295},
  month =     {June},
  address =   {Santa Clara, CA}}

@InProceedings{choi-dear-98,
  author =       {Jongmoo Choi and Sam H. Noh and Sang Lyul Min and and Yookun Cho},
  title =        {An adaptive block management scheme using on-line detection of block reference patterns},
  booktitle = {Proceedings of the 1998 International Workshop on Multimedia Database Management Systems (IW-MMDBMS 1998)},
  year =      1998,
  pages =     {172-179},
  month =     {August},
  address =   {Dayton, OH}}

@InProceedings{willick-eager-bunt-93,
  author =       {Darryl L. Willick and Derek L. Eager and Richard B. Bunt},
  title =        {Disk cache replacement policies for network fileservers},
  booktitle = {Proceedings of the 13th International Conference on Distributed Computing Systems(ICDCS 1993)},
  year =      1993,
  pages =     {2-11},
  month =     {May},
  address =   {Pittsburgh, PA}}

@MastersThesis{arlitt-thesis-96,
  author =       {Martin F. Arlitt},
  title =        {A performance study of Internet web servers},
  school =       {Department of Computer Science, University of Saskatchewan},
  year =         1996}

@PhdThesis{wenguang-thesis,
  author =       {Wenguang Wang},
  title =        {Storage Management for Large Scale Systems},
  school =       {Department of Computer Science, University of Saskatchewan},
  year =         2004,
  month =     {December}}

@ARTICLE{slru-94,
  author={Ramakrishna Karedla and Spencer Love and B. Wherry},
  journal={Computer},
  title={Caching strategies to improve disk system performance},
  year=1994,
  volume=27,
  number=3,
  pages={38-46},
  doi={10.1109/2.268884}}

@inproceedings{dlirs-li18,
  author       = {Cong Li},
  title        = {{DLIRS:} Improving Low Inter-Reference Recency Set Cache Replacement
                  Policy with Dynamics},
  booktitle    = {Proceedings of the 11th {ACM} International Systems and Storage Conference,
                  {SYSTOR} 2018, HAIFA, Israel, June 04-07, 2018},
  pages        = {59--64},
  publisher    = {{ACM}},
  year         = {2018},
  url          = {https://doi.org/10.1145/3211890.3211891},
  doi          = {10.1145/3211890.3211891},
  bibsource    = {dblp computer science bibliography, https://dblp.org}
}

@book{bovet2005linux,
  title={{Understanding the Linux Kernel}},
  author={Bovet, Daniel P. and Cesati, Marco},
  edition={3},
  year={2005},
  publisher={O'Reilly Media, Inc.}
}

@inproceedings {libCacheSim,
author = {Juncheng Yang and Yao Yue and K. V. Rashmi},
title = {A large scale analysis of hundreds of in-memory cache clusters at Twitter},
booktitle = {14th USENIX Symposium on Operating Systems Design and Implementation (OSDI 20)},
year = {2020},
isbn = {978-1-939133-19-9},
pages = {191--208},
url = {https://www.usenix.org/conference/osdi20/presentation/yang},
publisher = {USENIX Association},
month = nov,
}

@inproceedings{lru-Effelsberg1984,
  title={Principles of Database Buffer Management},
  author={Wolfgang Effelsberg and Theo Hrder},
  booktitle={ACM Transactions on Database Systems (TODS)},
  volume={9},
  issue={4},
  pages={560--595},
  year={1984},
  publisher={ACM}
}

@article{clock-Corbato1969,
  title={A Paging Experiment with the Multics System},
  author={Fernando J. Corbat and Victor A. Vyssotsky},
  journal={Honeywell Inc.},
  year={1969}
}

@inproceedings{car-Bansal2004,
author = {Bansal, Sorav and Modha, Dharmendra S.},
title = {CAR: Clock with Adaptive Replacement},
year = {2004},
publisher = {USENIX Association},
address = {USA},
booktitle = {Proceedings of the 3rd USENIX Conference on File and Storage Technologies},
pages = {187–200},
numpages = {14},
location = {San Francisco, CA},
series = {FAST '04}
}

@article{clockpro-Jiang2005,
  title={{Clock-Pro}: An Effective Improvement of the Clock Replacement},
  author={Song Jiang and Feng Chen and Xiaodong Zhang},
  journal={Proceedings of the USENIX Annual Technical Conference},
  pages={323--336},
  year={2005},
  organization={USENIX Association}
}

@inproceedings{lruk-O1993,
  title={{The LRU-K Page Replacement Algorithm For Database Disk Buffering}},
  author={Elizabeth J. O'Neil and Patrick E. O'Neil and Gerhard Weikum},
  booktitle={Proceedings of the 1993 ACM SIGMOD international conference on Management of data},
  pages={297--306},
  year={1993},
  organization={ACM}
}

@article{lfu-Sleator1985,
  title={Amortized Efficiency of List Update and Paging Rules},
  author={Daniel D. Sleator and Robert E. Tarjan},
  journal={Communications of the ACM},
  volume={28},
  number={2},
  pages={202--208},
  year={1985},
  publisher={ACM}
}

@article{lrfu-Lee1999,
  title={{LRFU}: A Spectrum of Policies that Subsumes the Least Recently Used and Least Frequently Used Policies},
  author={Donghee Lee and Jongmoo Choi and Jong-Hun Kim and Sam H. Noh and Sang Lyul Min and Yookun Cho and Chong Sang Kim},
  journal={IEEE Transactions on Computers},
  volume={50},
  number={12},
  pages={1352--1361},
  year={2001},
  publisher={IEEE}
}

@inproceedings{seq-glass97,
  title={Adaptive page replacement based on memory reference behavior},
  author={Gideon Glass and Pei Cao},
  booktitle={Proceedings of the 1997 ACM SIGMETRICS International Conference on Measurement and Modeling of Computer Systems (SIGMETRICS 1997)},
  pages={115--126},
  month={June},
  year={1997},
  organization={ACM},
  address={Seattle, WA, USA}
}

@inproceedings{eelru-99,
  title={{EELRU}: Simple and effective adaptive page replacement},
  author={Yannis Smaragdakis Scott Kaplan and Paul Wilson},
  booktitle={Proceedings of the 1999 ACM SIGMETRICS International Conference on Measurement and Modeling of Computer Systems (SIGMETRICS 1999)},
  pages={122--133},
  month=May,
  year={1999},
  organization={ACM},
  address={Atlanta, GA, USA}
}

@inproceedings{robinson90,
  title={Data cache management using frequency-based replacement},
  author={John T. Robinson and Murthy V. Devarakonda},
  booktitle={Proceedings of the 1990 ACM SIGMETRICS International Conference on Measurement and Modeling of Computer Systems (SIGMETRICS 1990)},
  pages={134--142},
  month={April},
  year={1990},
  organization={ACM},
  address={Boulder, CO, USA}
}

@inproceedings{lirs2-21,
author = {Zhong, Chen and Zhao, Xingsheng and Jiang, Song},
title = {LIRS2: an improved LIRS replacement algorithm},
year = {2021},
isbn = {9781450383981},
publisher = {Association for Computing Machinery},
address = {New York, NY, USA},
url = {https://doi.org/10.1145/3456727.3463772},
doi = {10.1145/3456727.3463772},
booktitle = {Proceedings of the 14th ACM International Conference on Systems and Storage},
articleno = {19},
numpages = {12},
location = {Haifa, Israel},
series = {SYSTOR '21}
}

@inproceedings{froese96,
  title={The effect of client caching on file server workloads},
  author={Kevin W. Froese and Richard B. Bunt},
  booktitle={Proceedings of the 29th Hawaii International Conference on System Sciences (HICSS 1996)},
  pages={150--159},
  month={January},
  year={1996},
  organization={ACM},
  address={Kihei, HI, USA}
}

@inproceedings{splinter-atc-20,
author = {Conway, Alex and Gupta, Abhishek and Chidambaran, Vijay and Farach-Colton, Martin and Spillane, Rick and Tai, Amy and Johnson, Rob},
title = {SplinterDB: closing the bandwidth gap for NVMe key-value stores},
year = {2020},
isbn = {978-1-939133-14-4},
publisher = {USENIX Association},
address = {USA},
booktitle = {Proceedings of the 2020 USENIX Conference on Usenix Annual Technical Conference},
articleno = {4},
numpages = {15},
series = {USENIX ATC'20}
}

@Misc{ext4tree,
  title =     {Understanding EXT4 (Part 3): Extent Trees},
  author = {Hal Pomeranz},
  howpublished = {\url{https://web.archive.org/web/20190818050155/https://digital-forensics.sans.org/blog/2011/03/28/digital-forensics-understanding-ext4-part-3-extent-trees}}}

@Misc{xfs,
  title = {{XFS}},
  author = {Wikipedia},
  howpublished = {\url{https://en.wikipedia.org/wiki/XFS}}
}

@misc{clock2q-patent,
  author       = {Wenguang Wang},
  title        = {System and methods of a CPU-efficient cache replacement algorithm},
  howpublished = {U.S. Patent 9760493B1},
  year         = {2017},
  month        = {September},
  note         = {Filed on {March 14, 2016}, issued on {September 12, 2017}},
}

@misc{clock2q-concurrent-patent,
  author       = {Wenguang Wang and Mounesh Badiger and Abhay Kumar Jain and Junlong Gao and Zhaohui Guo and Richard P. Spillane},
  title        = {System and method of a highly concurrent cache replacement algorithm},
  howpublished = {U.S. Patent 11086779B2},
  year         = {2021},
  month        = {August},
  note         = {Filed on {November 11, 2019}, issued on {August 10, 2021}},
}

@article{btrfs13,
  title={BTRFS: The Linux B-Tree Filesystem},
  author={Ohad Rodeh and Josef Bacik and Chris Mason},
  journal={ACM Transactions on Storage},
  volume={9},
  number={3},
  pages={1--32},
  year={2013},
  publisher={ACM}
}

@misc{cow-btree-code,
    title = {{COW B-trees}},
    author = {Ohad Rodeh},
    howpublished = {\url{https://github.com/ohad-rodeh/bt}}
}

@article{haerder1983principles,
  title={Principles of transaction-oriented database recovery},
  author={Haerder, Theo and Reuter, Andreas},
  journal={ACM Computing Surveys (CSUR)},
  volume={15},
  number={4},
  pages={287--317},
  year={1983},
  publisher={ACM}
}

@Misc{TLX,
  title = 	 {{TLX}: Collection of Sophisticated {C++} Data Structures, Algorithms, and Miscellaneous Helpers},
  author = 	 {Timo Bingmann},
  year = 	 2018,
  note = 	 {\url{https://panthema.net/tlx}, retrieved {Oct.} 7, 2020},
}

@inproceedings {3lcache,
author = {Wenbin Zhou and Zhixiong Niu and Yongqiang Xiong and Juan Fang and Qian Wang},
title = {{3L-Cache}: Low Overhead and Precise Learning-based Eviction Policy for Caches},
booktitle = {23rd USENIX Conference on File and Storage Technologies (FAST 25)},
year = {2025},
isbn = {978-1-939133-45-8},
address = {Santa Clara, CA},
pages = {237--254},
url = {https://www.usenix.org/conference/fast25/presentation/zhou-wenbin},
publisher = {USENIX Association},
month = feb
}

@inproceedings{zhang_sieve_2024,
	title = {{SIEVE} is {Simpler} than {LRU}: an {Efficient} {Turn}-{Key} {Eviction} {Algorithm} for {Web} {Caches}},
	isbn = {978-1-939133-39-7},
	shorttitle = {{SIEVE} is {Simpler} than \{{LRU}\}},
	language = {en},
	urldate = {2024-07-21},
	booktitle = {21st {USENIX} {Symposium} on {Networked} {Systems} {Design} and {Implementation} ({NSDI} 24)},
	author = {Zhang, Yazhuo and Yang, Juncheng and Yue, Yao and Vigfusson, Ymir and Rashmi, K. V.},
	year = {2024},
	pages = {1229--1246},
}

@inproceedings{yang_fifo_2023,
	address = {New York, NY, USA},
	series = {{HOTOS}'23},
	title = {{FIFO} can be {Better} than {LRU}: the {Power} of {Lazy} {Promotion} and {Quick} {Demotion}},
	isbn = {9798400701955},
	shorttitle = {{FIFO} can be {Better} than {LRU}},
	doi = {10.1145/3593856.3595887},
	urldate = {2024-01-25},
	booktitle = {Proceedings of the 19th {Workshop} on {Hot} {Topics} in {Operating} {Systems}},
	publisher = {Association for Computing Machinery},
	author = {Yang, Juncheng and Qiu, Ziyue and Zhang, Yazhuo and Yue, Yao and Rashmi, K. V.},
	month = jun,
	year = {2023},
	pages = {70--79},
}

@inproceedings{yang_fifo_2023-1,
	address = {New York, NY, USA},
	series = {{SOSP}'23},
	title = {{FIFO} queues are all you need for cache eviction},
	isbn = {9798400702297},
	doi = {10.1145/3600006.3613147},
	urldate = {2024-01-25},
	booktitle = {Proceedings of the 29th {Symposium} on {Operating} {Systems} {Principles}},
	publisher = {Association for Computing Machinery},
	author = {Yang, Juncheng and Zhang, Yazhuo and Qiu, Ziyue and Yue, Yao and Vinayak, Rashmi},
	month = oct,
	year = {2023},
	pages = {130--149},
}

@inproceedings{song_halp_2023,
	series = {{NSDI}'23},
	title = {{HALP}: {Heuristic} {Aided} {Learned} {Preference} {Eviction} {Policy} for {YouTube} {Content} {Delivery} {Network}},
	isbn = {978-1-939133-33-5},
	shorttitle = {{HALP}},
	language = {en},
	urldate = {2023-04-19},
	booktitle = {20th {USENIX} {Symposium} on {Networked} {Systems} {Design} and {Implementation}},
	author = {Song, Zhenyu and Chen, Kevin and Sarda, Nikhil and Altinbuken, Deniz and Brevdo, Eugene and Coleman, Jimmy and Ju, Xiao and Jurczyk, Pawel and Schooler, Richard and Gummadi, Ramki},
	year = {2023},
	pages = {1149--1163},
}

@inproceedings{yang_gl-cache_2023,
	series = {{FAST}'23},
	title = {{GL}-{Cache}: {Group}-level learning for efficient and high-performance caching},
	isbn = {978-1-939133-32-8},
	shorttitle = {{GL}-{Cache}},
	language = {en},
	urldate = {2023-04-17},
	booktitle = {21st {USENIX} {Conference} on {File} and {Storage} {Technologies} ({FAST} 23)},
	author = {Yang, Juncheng and Mao, Ziming and Yue, Yao and Rashmi, K. V.},
	year = {2023},
	pages = {115--134},
}

@inproceedings{rodriguez_learning_2021,
	series = {{FAST}'21},
	title = {Learning {Cache} {Replacement} with {CACHEUS}},
	isbn = {978-1-939133-20-5},
	shorttitle = {cacheus},
	booktitle = {19th {USENIX} {Conference} on {File} and {Storage} {Technologies}},
	publisher = {USENIX Association},
	author = {Rodriguez, Liana V. and Yusuf, Farzana and Lyons, Steven and Paz, Eysler and Rangaswami, Raju and Liu, Jason and Zhao, Ming and Narasimhan, Giri},
	month = feb,
	year = {2021},
	pages = {341--354},
}

@inproceedings{vietri_driving_2018,
	address = {Boston, MA},
	series = {{hotStorage}'18},
	title = {Driving cache replacement with {ML}-based {LeCaR}},
	shorttitle = {{LeCaR}},
	booktitle = {10th {USENIX} workshop on hot topics in storage and file systems},
	publisher = {USENIX Association},
	author = {Vietri, Giuseppe and Rodriguez, Liana V. and Martinez, Wendy A. and Lyons, Steven and Liu, Jason and Rangaswami, Raju and Zhao, Ming and Narasimhan, Giri},
	month = jul,
	year = {2018},
}

@inproceedings{blankstein_hyperbolic_2017,
	address = {Santa Clara, CA},
	series = {{ATC}'17},
	title = {Hyperbolic caching: {Flexible} caching for web applications},
	isbn = {978-1-931971-38-6},
	shorttitle = {Hyperbolic},
	booktitle = {2017 {USENIX} annual technical conference},
	publisher = {USENIX Association},
	author = {Blankstein, Aaron and Sen, Siddhartha and Freedman, Michael J.},
	month = jul,
	year = {2017},
	pages = {499--511},
}

@article{einziger_tinylfu_2017,
	series = {{TOS}'17},
	title = {{TinyLFU}: {A} {Highly} {Efficient} {Cache} {Admission} {Policy}},
	volume = {13},
	issn = {1553-3077, 1553-3093},
	shorttitle = {{TinyLFU}},
	doi = {10.1145/3149371},
	language = {en},
	number = {4},
	urldate = {2022-07-18},
	journal = {ACM Transactions on Storage},
	author = {Einziger, Gil and Friedman, Roy and Manes, Ben},
	month = dec,
	year = {2017},
	pages = {1--31},
}

@inproceedings{waldspurger_efficient_2015,
	address = {Santa Clara, CA},
	series = {{FAST}'15},
	title = {Efficient {MRC} construction with {SHARDS}},
	isbn = {978-1-931971-20-1},
	shorttitle = {{SHARDS}},
	booktitle = {13th {USENIX} conference on file and storage technologies},
	publisher = {USENIX Association},
	author = {Waldspurger, Carl A. and Park, Nohhyun and Garthwaite, Alexander and Ahmad, Irfan},
	month = feb,
	year = {2015},
	pages = {95--110},
}

@inproceedings{huang_analysis_2013,
	address = {New York, NY, USA},
	series = {{SOSP}'13},
	title = {An analysis of {Facebook} photo caching},
	isbn = {978-1-4503-2388-8},
	shorttitle = {{SLRU}},
	doi = {10.1145/2517349.2522722},
	urldate = {2023-01-17},
	booktitle = {Proceedings of the {Twenty}-{Fourth} {ACM} {Symposium} on {Operating} {Systems} {Principles}},
	publisher = {Association for Computing Machinery},
	author = {Huang, Qi and Birman, Ken and van Renesse, Robbert and Lloyd, Wyatt and Kumar, Sanjeev and Li, Harry C.},
	month = nov,
	year = {2013},
	pages = {167--181},
}

@article{yadgar_management_2011,
	title = {Management of {Multilevel}, {Multiclient} {Cache} {Hierarchies} with {Application} {Hints}},
	volume = {29},
	issn = {0734-2071},
	doi = {10.1145/1963559.1963561},
	number = {2},
	urldate = {2023-08-01},
	journal = {ACM Transactions on Computer Systems},
	author = {Yadgar, Gala and Factor, Michael and Li, Kai and Schuster, Assaf},
	month = may,
	year = {2011},
	pages = {5:1--5:51},
}

@inproceedings{yadgar_karma_2007,
	series = {{FAST}'07},
	title = {Karma: {Know}-it-{All} {Replacement} for a {Multilevel} {Cache}},
	shorttitle = {Karma},
	language = {en},
	urldate = {2023-08-01},
	booktitle = {5th {USENIX} {Conference} on {File} and {Storage} {Technologies} ({FAST} 07)},
	author = {Yadgar, Gala and Factor, Michael and Schuster, Assaf},
	year = {2007},
}

@inproceedings{jiang_lirs_2002,
	series = {{SIGMETRICS}'02},
	title = {{LIRS}: an efficient low inter-reference recency set replacement policy to improve buffer cache performance},
	volume = {30},
	shorttitle = {{LIRS}},
	doi = {10.1145/511399.511340},
	urldate = {2023-01-06},
	booktitle = {{ACM} {SIGMETRICS} {Performance} {Evaluation} {Review}},
	author = {Jiang, Song and Zhang, Xiaodong},
	month = jun,
	year = {2002},
	pages = {31--42},
}

@inproceedings{carr_wsclock_1981,
	address = {New York, NY, USA},
	series = {{SOSP}'81},
	title = {{WSCLOCK}: a simple and effective algorithm for virtual memory management},
	isbn = {978-0-89791-062-0},
	doi = {10.1145/800216.806596},
	urldate = {2023-02-02},
	booktitle = {Proceedings of the eighth {ACM} symposium on {Operating} systems principles},
	publisher = {Association for Computing Machinery},
	author = {Carr, Richard W. and Hennessy, John L.},
	month = dec,
	year = {1981},
	pages = {87--95},
}

@book{cherkasova_improving_1998,
	title = {Improving {WWW} proxies performance with greedy-dual-size-frequency caching policy},
	shorttitle = {{GDSF}},
	publisher = {Citeseer},
	author = {Cherkasova, Ludmila},
	year = {1998},
}

@inproceedings{johnson_2q_1994,
	address = {San Francisco, CA, USA},
	series = {{VLDB}'94},
	title = {{2Q}: {A} {Low} {Overhead} {High} {Performance} {Buffer} {Management} {Replacement} {Algorithm}},
	isbn = {978-1-55860-153-6},
	shorttitle = {{2Q}},
	urldate = {2023-01-06},
	booktitle = {Proceedings of the 20th {International} {Conference} on {Very} {Large} {Data} {Bases}},
	publisher = {Morgan Kaufmann Publishers Inc.},
	author = {Johnson, Theodore and Shasha, Dennis},
	month = sep,
	year = {1994},
	pages = {439--450},
}

@inproceedings{megiddo_arc_2003,
	series = {{FAST}'03},
	title = {{ARC}: {A} self-tuning, low overhead replacement cache},
	shorttitle = {{ARC}},
	booktitle = {2nd {USENIX} conference on file and storage technologies},
	author = {Megiddo, Nimrod and Modha, Dharmendra S},
	year = {2003},
}

@inproceedings{tang_ripq_2015,
	series = {{FAST}'15},
	title = {{RIPQ}: {Advanced} photo caching on flash for facebook},
	shorttitle = {{RIPQ}},
	booktitle = {13th {USENIX} {Conference} on {File} and {Storage} {Technologies}},
	author = {Tang, Linpeng and Huang, Qi and Lloyd, Wyatt and Kumar, Sanjeev and Li, Kai},
	year = {2015},
	pages = {373--386},
}

@inproceedings{beckmann_lhd_2018,
	series = {{NSDI}'18},
	title = {{LHD}: {Improving} cache hit rate by maximizing hit density},
	shorttitle = {{LHD}},
	booktitle = {15th {USENIX} symposium on networked systems design and implementation},
	author = {Beckmann, Nathan and Chen, Haoxian and Cidon, Asaf},
	year = {2018},
	pages = {389--403},
}

@inproceedings{song_learning_2020,
	series = {{NSDI}'20},
	title = {Learning relaxed belady for content distribution network caching},
	shorttitle = {{LRB}},
	booktitle = {17th {USENIX} symposium on networked systems design and implementation},
	author = {Song, Zhenyu and Berger, Daniel S and Li, Kai and Shaikh, Anees and Lloyd, Wyatt and Ghorbani, Soudeh and Kim, Changhoon and Akella, Aditya and Krishnamurthy, Arvind and Witchel, Emmett and {others}},
	year = {2020},
	pages = {529--544},
}
\bibliographystyle{ACM-Reference-Format}

\end{document}